\documentclass{ieeetj}
\usepackage{cite}
\usepackage{amsmath,amssymb,amsfonts}
\usepackage{algorithmic}
\usepackage{graphicx,color}
\usepackage{textcomp}
\usepackage{xcolor}
\usepackage{hyperref}
\usepackage{siunitx}
\usepackage{subfigure}
\hypersetup{hidelinks=true}
\usepackage{algorithm,algorithmic}
\def\BibTeX{{\rm B\kern-.05em{\sc i\kern-.025em b}\kern-.08em
    T\kern-.1667em\lower.7ex\hbox{E}\kern-.125emX}}
\AtBeginDocument{\definecolor{tmlcncolor}{cmyk}{0.93,0.59,0.15,0.02}\definecolor{NavyBlue}{RGB}{0,86,125}}

\def\OJlogo{\vspace{-4pt}$<$Society logo(s) and publication title will appear here.$>$}
\def\seclogo{\vspace{10pt}$<$Society logo(s) and publication title will appear here.$>$}

\def\authorrefmark#1{\ensuremath{^{\textbf{#1}}}}

\begin{document}
\receiveddate{XX Month, XXXX}
\reviseddate{XX Month, XXXX}
\accepteddate{XX Month, XXXX}
\publisheddate{XX Month, XXXX}
\currentdate{XX Month, XXXX}
\doiinfo{XXXX.2022.1234567}

\markboth{}{Thornton {et al.}}

\title{Decoding Envelope and Frequency-Following EEG Responses to Continuous Speech Using Deep Neural Networks}

\author{Mike Thornton\authorrefmark{1}, Danilo P. Mandic\authorrefmark{2}, Fellow, IEEE,\\ and Tobias  Reichenbach\authorrefmark{3}}
\affil{Department of Computing, Imperial College London, SW7 2RH, U.K.}
\affil{Department of Electrical and Electronic Engineering, Imperial College London, SW7 2RH, U.K.}
\affil{Department for Artificial Intelligence in Biomedical Engineering, Friedrich-Alexander-University Erlangen-Nürnberg, 91052 Erlangen, Germany}
\corresp{Corresponding author: Tobias Reichenbach (email: tobias.j.reichenbach@fau.de).}
\authornote{Mike Thornton is supported by the UKRI CDT in AI for Healthcare \href{http://ai4health.io}{http://ai4health.io} (Grant No. P/S023283/1)}

\begin{abstract}
The electroencephalogram (EEG) offers a non-invasive means by which a listener's auditory system may be monitored during continuous speech perception. Reliable auditory-EEG decoders could facilitate the objective diagnosis of hearing disorders, or find applications in cognitively-steered hearing aids. Previously, we developed decoders for the ICASSP Auditory EEG Signal Processing Grand Challenge (SPGC). These decoders aimed to solve the match-mismatch task: given a short temporal segment of EEG recordings, and two candidate speech segments, the task is to identify which of the two speech segments is temporally aligned, or matched, with the EEG segment. The decoders made use of cortical responses to the speech envelope, as well as speech-related frequency-following responses, to relate the EEG recordings to the speech stimuli. Here we comprehensively document the methods by which the decoders were developed. We extend our previous analysis by exploring the association between speaker characteristics (pitch and sex) and classification accuracy, and provide a full statistical analysis of the final performance of the decoders as evaluated on a heldout portion of the dataset. Finally, the generalisation capabilities of the decoders are characterised, by evaluating them using an entirely different dataset which contains EEG recorded under a variety of speech-listening conditions. The results show that the match-mismatch decoders achieve accurate and robust classification accuracies, and they can even serve as auditory attention decoders without additional training.
\end{abstract}

\begin{IEEEkeywords}
Auditory attention decoding, Deep learning, EEG signal processing
\end{IEEEkeywords}


\maketitle

\section{INTRODUCTION}
\IEEEPARstart{T}{he} neural processes by which normal-hearing human listeners perceive and understand spoken language are not well understood. These processes must be executed rapidly enough for listeners to be able to comprehend speech in real time, and resilient enough to preserve speech comprehension under adverse listening conditions. For example, during selective attention to one of several speech streams, the auditory system performs processing to enhance the attended speech stream: this is the famous cocktail-party effect~\cite{cherry53}. 

The auditory system produces electric and magnetic signals during speech perception. These signals may be recorded noninvasively and at high sampling rates, offering the opportunity to study the rapid nature of continuous speech processing. Such signals are notoriously noisy, due to artefacts originating from the activity of participants' hearts, eyes, and muscles, as well as external electromagnetic fields. Moreover, noninvasive recordings of electric and magnetic signals are inevitably contaminated by background neural activity which is not related to auditory processing. A significant challenge therefore lies in identifying and isolating auditory contributions to the recorded signals.

Techniques for system identification have facilitated much of the recent research into electroencephalography (EEG) and magnetoencephalography (MEG) responses to continuous speech~\cite{Lalor2010}. The temporal response function (TRF) approach aims to identify the linear time-invariant system that best describes M/EEG responses to features of the stimulating speech stream~\cite{Lalor2010, Ding2012, brodbeck_continuous_2020}. This approach has proven particularly fruitful for characterising M/EEG responses to the temporal envelope of speech, a feature which primarily captures slow amplitude fluctuations driven by words and syllables. A number of studies have now demonstrated that cortical responses to the speech envelope are modulated by selective auditory attention~\cite{Ding2012attention, OSullivan2014}. Smart hearing aids which leverage this effect could one day improve outcomes for hearing-impaired listeners, by selectively amplifying particular sources in busy auditory scenes according to the focus of the user's auditory attention~\cite{geirnaert2020neurosteered}.

Beyond temporal envelope processing, the TRF method has been used to explore the neural correlates of a wide range of cognitive and perceptual factors of speech, including linguistic surprisal, speech-in-noise clarity, and speech comprehension levels, to name but a few examples~\cite{Etard2019, Lesenfants2019, Weissbart2020, Broderick2021}. Recently, responses which phase-lock to the fundamental waveform of continuous speech (F0), as well as to the envelope modulations of its higher harmonics, have also attracted considerable interest~\cite{forte2017human, Hertrich2011, Kulasingham2020}. These responses are termed speech-related frequency-following responses, or speech-FFRs. Classical frequency-following responses are evoked by simple stimuli such as tones or vowels, and are detected by averaging the M/EEG recordings over many repetitions of the stimulus sound. Aiken and Picton distinguish between two types of FFR: the spectral FFRs, which phase-lock to spectral components of the stimulus (such as the harmonics of a vowel sound) that are resolved by the cochlea; and the envelope FFR, which phase-locks to high-frequency periodicity in the envelope of the stimulus (e.g. the glottal pitch envelope of a vowel sound)~\cite{Aiken2008}. Speech-FFRs which phase-lock to F0 can be considered direct analogues of the spectral FFR, and speech-FFRs which phase-lock to the high-frequency envelope modulations of the speech waveform can be considered analogous to the envelope FFR.

Speech-FFR waveforms can be obtained through TRFs (i.e. deconvolution), rather than through averaging. Both show strong responses at the fundamental frequency of speech, with envelope-related speech-FFRs exhibiting a much stronger amplitude than spectral speech-FFRs~\cite{Kegler2022, Kulasingham2020}. The speech-FFRs have also been shown to be modulated by selective attention to speech, with attended voices eliciting stronger responses than unattended voices~\cite{brainstemDecoding, Schller2023}. Speech-FFRs are also affected by other speaker characteristics, particularly pitch, but also the speech rate and the variability of the speaker's pitch~\cite{Hertrich2011, VanCanneyt2021, Kulasingham2020}.

As an alternative to TRFs (which are linear models), nonlinear methods such as deep neural networks (DNNs) may be used to relate M/EEG recordings to continuous speech. The literature concerning the application of DNNs for auditory EEG decoding is growing particularly quickly~\cite{puffay2023relating}. Several factors motivate the use of DNNs for decoding EEG responses to speech. First, DNNs are better suited than TRFs to capture the inherently nonlinear nature of the auditory system. Second, the decoding performance of DNNs can reflect perceptual factors such as speech-in-noise intelligibility, potentially facilitating the objective diagnosis of hearing disorders, or the objective evaluation of listening devices~\cite{Accou2021}. Finally, DNNs have been shown to possess a remarkable ability to generalise across individuals, even though the characteristics of EEG signals are highly individual-specific~\cite{Accou2021, Accou2023}. A single linear model cannot usually be used to accurately decode EEG signals recorded from a cohort of individuals. For many applications, including cognitively-steered hearing aids, a decoder which works for unseen individuals in an `out-of-the-box' fashion would be highly desirable.

\begin{figure*}[ht!]
    \centering

    \subfigure[]{\includegraphics[width=0.49\textwidth]{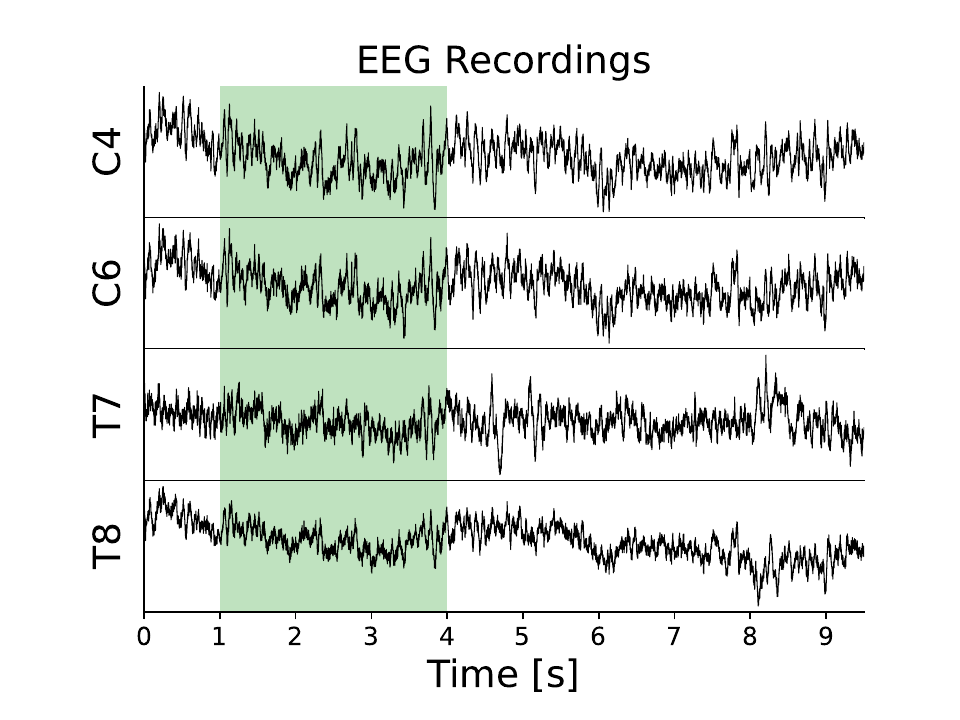}}\hfill
    \subfigure[]{\includegraphics[width=0.49\textwidth]{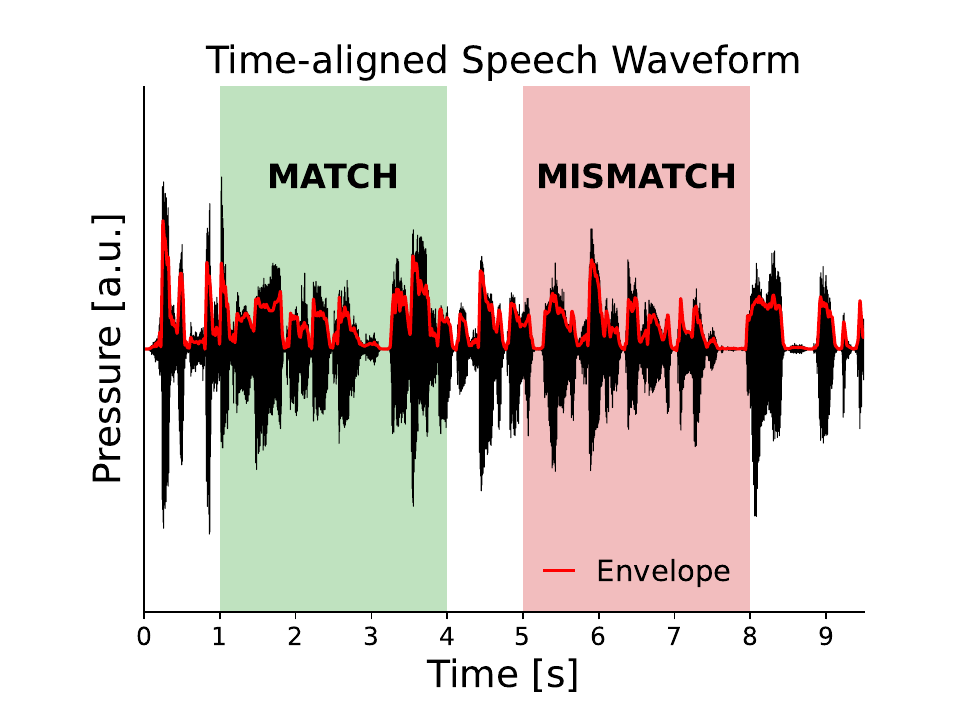}}
    \caption{Overview of the match-mismatch task. a) Multi-channel time-series of EEG recordings over a ten-second period. A short temporal segment of the EEG signals is highlighted in green. b) The time-aligned speech waveform is shown on the right. A segment which is time-aligned, or matched, with the EEG segment is highlighted in green. A mismatched segment is highlighted in red. In the match-mismatch task, an auditory EEG decoder must learn to distinguish between matched and mismatched stimulus segments, given a short temporal segment of EEG recordings.}    \label{fig:match-mismatch-overview}
\end{figure*}

In this work, we present and further develop our deep-learning approach for decoding EEG responses to speech, originally designed for the ICASSP 2023 Auditory EEG Decoding Signal Processing Grand Challenge (SPGC)~\cite{jalilpour2022icassp, ICASSPTHORNTON}. We focus on the match-mismatch sub-task of the challenge: given a short temporal segment of multichannel EEG, as well as two candidate speech segments, the task is to identify which of the two speech segments is temporally aligned with the EEG segment (see Figure~\ref{fig:match-mismatch-overview}). This auditory match-mismatch paradigm was originally proposed by de Cheveigne~\textit{et al.} and is free from a number of potential confounds which are present in the more common auditory attention decoding paradigm; for example, two-talker selective-attention tasks are more cognitively demanding for participants than are single-talker active listening tasks, which could lead to decreased rates of compliance with the experimental protocol~\cite{deCheveign2021JNE}. Subsequent work has shown that both acoustic and linguistic features derived from speech signals carry information which can be decoded from EEG signals using deep neural networks in a match-mismatch paradigm. These features include the temporal envelope of speech, the mel spectrogram, phonetic features, and word-level features such as word frequency and word surprisal~\cite{monesi2021interspeech, Accou2021EUSIPCO, Puffay2023JNEb, Soman2023Interspeech}.

The match-mismatch decoders developed in this work were evaluated against data from `seen' participants, who already featured in the training dataset, as well as data recorded from `unseen' participants. Our approach to the match-mismatch decoding problem is similar to that of Puffay~\textit{et al.} in that we sought to exploit cortical tracking of the speech temporal envelope as well as a speech-FFR through deep neural networks~\cite{puffay22_interspeech}. However, those authors made use of the spectral speech-FFR, whereas we decided to use the envelope-related speech-FFR, which has a stronger magnitude at the fundamental frequency of speech~\cite{Kegler2022, Kulasingham2020}. Two further approaches for boosting the accuracy of the decoding system are explored in this work: fine-tuning of the decoders to individual listeners, and ensembling a population of distinct decoders by averaging over their individual predictions. Finally, a comprehensive assessment of the generalisation capabilities of the decoders is provided, including an investigation into their application as auditory attention decoders.

\section{Materials and Methods}

\subsection{Datasets}

The authors of the ICASSP 2023 Auditory EEG Decoding SPGC provided a very large dataset comprising EEG measurements taken from participants who listened to speech material in their native language (Dutch)~\cite{K3VSND_2023}. This dataset is named SparrKULee. A total of 85 young and normal-hearing participants are included in this dataset. The time-aligned speech material, which consists of audiobooks and podcasts, is also included in the dataset. The speech material was presented by both male and female speakers. Each participant underwent between four and twelve trials, and one entire audiobook chapter or podcast was presented per trial (average trial duration: about fifteen minutes.)

SparrKULee was divided into two parts, termed the development dataset and the heldout dataset. We used the development dataset to experiment with our decoders, and the accuracy of the final decoding system was evaluated using the heldout dataset. The development dataset consists of EEG recordings from 71 participants, and between three and eleven trials per participant. Each trial in the development dataset was further split into three non-overlapping portions: the training portion, consisting of the first \qty{80}{\percent} of the recorded data; the validation portion, consisting of the next \qty{10}{\percent}; and the testing portion, which consists of the remaining \qty{10}{\percent} of the data at the end of the trial. Overall, the three portions of the development dataset contained EEG recorded for 94.13 hours, 11.77 hours, and 11.77 hours, respectively. The heldout dataset contained EEG recordings from the same 71 participants whilst they listened to distinct speech material, as well as EEG recordings from an additional 14 unseen participants. The heldout dataset contained between one and seven trials per participant, totalling almost 40 listening hours (seen participants: 19.61 hours; unseen participants: 19.34 hours). 

The EEG recordings were acquired from 64 channels at a sampling rate of \qty{8196}{\hertz} using a Biosemi ActiveTwo system (Biosemi, Netherlands). The electrodes were applied to the scalp of each participant with conductive gel, and positioned according to the international 10-20 system using the Biosemi 64-channel electrode cap (Biosemi, Netherlands). Although the recordings were acquired at \qty{8196}{\hertz}, the public version of the dataset provides resampled recordings with a sampling rate of \qty{1024}{\hertz}. The resampled EEG recordings are therefore used in the present work.

We applied our trained decoders to a second publicly-available EEG dataset, which will be referred to as the ICL dataset~\cite{EtardDataset}. Eighteen native-English-speaking participants listened to audiobooks presented both in quiet and noisy conditions. The noisy conditions include speech presented in three levels of background babble noise (with signal-to-noise ratios (SNRs) of \qty{0.4}{\dB}, \qty{-1.4}{\dB}, and \qty{-3.2}{\dB}), as well as two competing-speakers conditions. In the first, each listener was asked to attend to a male narrator whilst ignoring a female narrator. In the second condition, the roles of the two speakers (attended vs ignored) were swapped. Furthermore, in a separate recording session, twelve participants listened to audiobooks which were presented in a foreign language that they did not understand (Dutch). Of these participants, ten had already taken part in the English recording session. During the Dutch session, the participants listened to speech in quiet conditions, as well as in the same three noisy conditions described above; however, we only used the foreign-language-in-quiet condition in this work. Therefore, we applied our already-trained decoders in seven distinct listening conditions. Each listening condition was split into four trials, each of approximately \qty{2.5}{minutes} in duration. The EEG signals were acquired at a sampling rate of \qty{1000}{\hertz} via an ActiCHamp amplifier (BrainProducts, Germany) and 63 active electrodes applied to the scalp with conductive gel. The electrodes were positioned according to the easycap-M1 electrode cap (BrainProducts, Germany), which conforms to the international 10-20 system. An additional electrode was placed on the right earlobe and served as the common reference electrode. Therefore, this dataset contained 63-channel EEG recordings.

\subsection{Data pre-processing}

\subsubsection{SparrKULee}

Representations of the speech envelopes were calculated according to a procedure inspired by the auditory system, as described by Biesmans \textit{et al.}~\cite{Biesmans2017}. Each raw speech waveform (sampled at 48 kHz) was first passed through a gammatone filterbank composed of 28 filters spaced equidistantly on an ERB (equivalent rectangular bandwidth) scale between \qty{50}{\hertz} and \qty{5}{kHz}. The resulting sub-band waveforms were subsequently full-wave rectified and raised to the power of 0.6 in order to obtain compressed sub-band envelopes. These were averaged to produce a single-channel envelope, which was finally resampled to \qty{64}{\hertz}.

Following previous studies, we used the high-frequency envelope modulations feature to represent periodicity in the speech envelopes~\cite{Hertrich2011, Kulasingham2020}. The speech waveforms were first resampled to \qty{16}{kHz}. A time-frequency representation (auditory spectrogram) of each speech waveform’s power was then obtained from a publicly-available biophysical model of the auditory periphery~\cite{NSL, KeglerPyNSL}. The auditory spectrograms have a reduced sampling rate of \qty{500}{\hertz}. The frequency bins corresponding to centre frequencies between \qty{300}{\hertz} and \qty{4}{kHz} were averaged, and the subsequent waveform was bandpass filtered in the range of \qtyrange{70}{220}{\hertz}, which is approximately the range of the fundamental frequency of speech (FIR sinc-Hamming functions of order 249 applied twice via forward and backward passes). The resulting signal was resampled to \qty{512}{\hertz}.

We produced a pre-processed version of the EEG recordings from SparrKULee to accompany the speech envelopes using the \textit{brain\_pipe} package~\cite{brainpipeNEW}. The pre-processing pipeline was developed by the organisers of the SPGC. First, slow drifts were removed from the EEG recordings using a highpass infinite impulse response (IIR) filter (first-order Butterworth filter with \qty{-3}{dB} attenuation at \qty{0.5}{\hertz}, applied twice via forward and backward passes). Then, a simple single-channel threshold-and-interpolate procedure was employed to identify and remove noisy segments in the recordings. Events with an amplitude exceeding \qty{500}{\si{\micro\volt}} were marked as glitches. Glitchy segments were then replaced using linear interpolation, where the values were derived by interpolating between the samples immediately before and after each glitchy segment.

A second threshold-based artifact suppression routine was then applied. In this case, the five most frontal channels (the Fp/AF channels, as well as Fz) were averaged to form a mean frontal channel. Next, the average power of this mean frontal channel over the course of each trial was calculated and used to define a threshold for identifying glitchy segments. Specifically, the threshold power was defined to be five times the average power of the mean frontal channel; any time instances at which the instantaneous power of the mean frontal channel exceeded this threshold power were labelled as glitchy. For each trial, a multichannel Wiener filter was fitted by using the segments marked as `clean' and `glitchy' to estimate the covariance matrices of both the clean EEG signals as well the (assumed-to-be) superposed artefact signals, respectively. The filter was then applied to the entire EEG signal in order to suppress high-power artefacts, for example due to blinks or movements~\cite{Somers2018}. The EEG recordings were re-referenced to the average voltage of all the electrodes and finally resampled to the same sampling frequency as the speech envelopes (\qty{64}{\hertz}.)

A second pre-processed version of the same EEG recordings was produced to accompany the high-frequency envelope modulations feature. The EEG recordings were detrended via the same high-pass IIR filter described above, and the same initial threshold-and-interpolate procedure was then applied. Next, the EEG recordings were average-referenced, bandpass filtered between \qtyrange{70}{220}{\hertz} (FIR type-I sinc-Hamming functions with a duration of \qty{1}{\second} applied twice via forward and backward passes), and resampled to \qty{512}{\hertz} (the sampling rate of the high-frequency envelope-modulations feature.)

\subsubsection{ICL dataset}
The speech envelopes and the high-frequency envelope modulations features were extracted from the ICL audiobooks using the same procedures described above. The EEG pre-processing pipelines were also similar, except that differences in the electrode layouts of the two datasets needed to be accounted for. After applying the aforementioned filtering and artefact-suppression routines to the ICL EEG signals, five channels (Fpz, Iz, P9, P10, PO4) which were used in SparrKULee but missing from the ICL dataset were interpolated. Then, four channels (FT9, FT10, TP9, TP10) which were not used in SparrKULee were dropped. The common-average EEG re-referencing procedure was applied after this step.

\subsection{Deep neural networks}
\label{sec:methods-dnns}
\begin{figure*}[ht!]
\centering
    \includegraphics[width=1\linewidth]{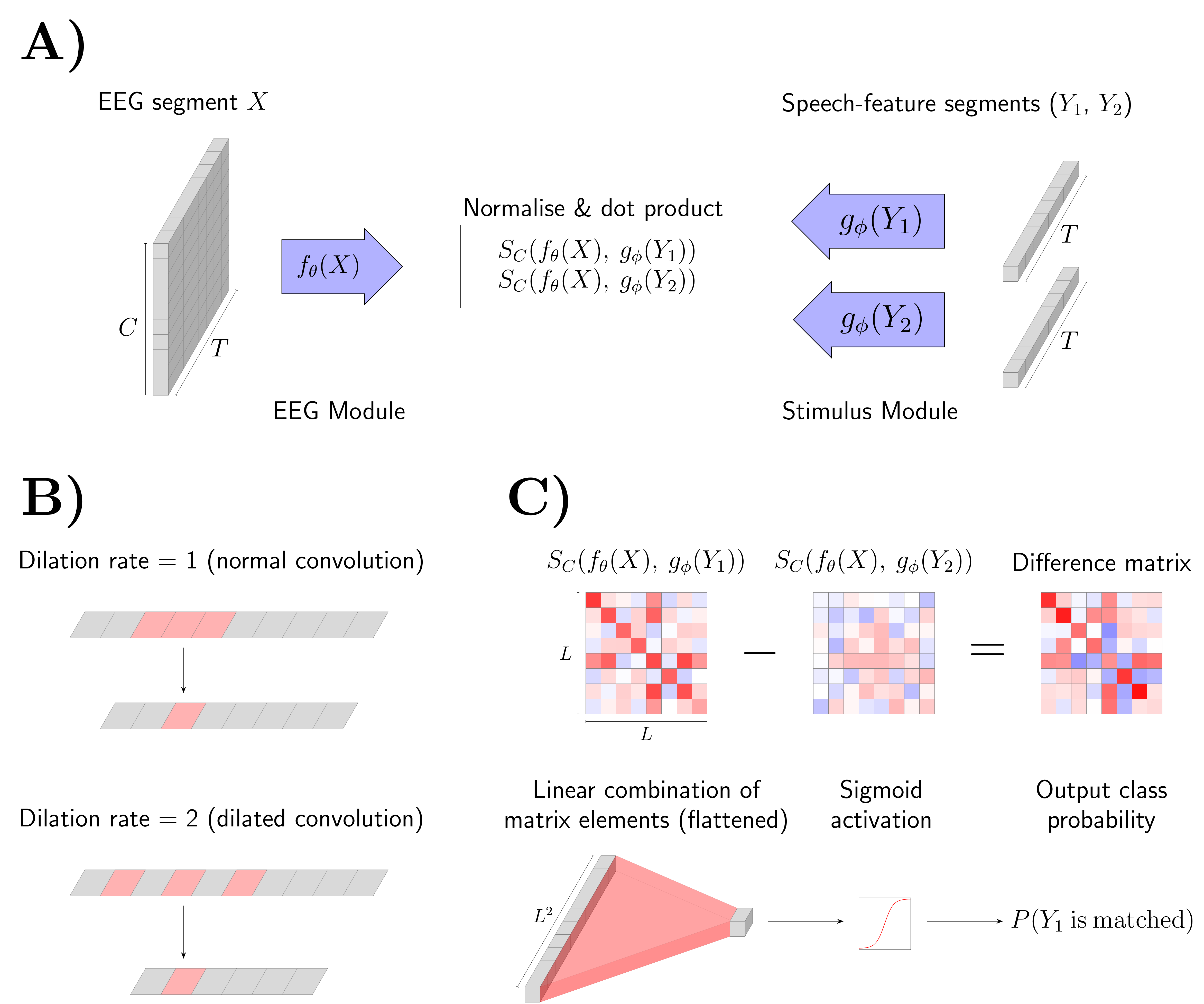}
    \caption{High-level overview of the deep neural network architecture used in this work. A) The operation of the EEG module and the stimulus module. The EEG module accepts a three-second temporal segment of 64-channel EEG recordings as an input, to which a nonlinear transformation $f_\theta$ is subsequently applied. The stimulus module accepts a univariate three-second segment of a feature of the stimulus (i.e. the envelope modulations, or the speech envelope), to which it applies a nonlinear transformation $g_\phi$. The EEG and stimulus segments are each transformed into timeseries with $L=16$ channels, which are compared through the cosine similarity metric $S_C$. B) Demonstration of the action of dilated convolutions. Dilated convolutions employ spaces between the convolutional kernel elements, widening the receptive field of the convolutional layer. Since no padding is applied, the output channel contains less temporal samples than the input channel(s), as shown. C) Overview of the output layer of the decoder. Each channel of the transformed EEG segment is compared against each channel of the transformed stimulus segment through the cosine similarity metric. This results in an ordered pair $16\times16$ matrices, (one matrix for each of the two candidate stimulus segments). The matrices are subtracted, and a linear combination of the resulting 256 matrix elements is taken to produce a scalar logit. The probability that the first stimulus segment $Y_1$ is the matched segment is calculated by applying the sigmoid function to this logit.}
    \label{fig:architecture}
\end{figure*}

The deep learning architecture used in this work is based on the deep neural network (DNN) architecture originally proposed by Accou~\textit{et al}.~\cite{Accou2021EUSIPCO}. It consists of two modules - an EEG module, and a stimulus module. These two modules respectively project the EEG and stimulus segments into a space where matched segments are maximally similar. This process is shown diagrammatically in Figure~\ref{fig:architecture}a. The stimulus segments are represented by segments of the features of the speech streams (the envelopes or the high-frequency envelope modulations feature). Both modules employ one-dimensional convolutional layers. A convolutional layer implements a set of multi-channel matched filters of a fixed length (or kernel size)~\cite{mandicmatchedfilters, Stankovic2023SPM}. The number of output channels of a convolutional layer is determined by the number of matched filters that the layer implements. Importantly, each matched filter is parameterised by a matrix of learnable weights (of shape $C\times K$, where $C$ is the number of input channels and $K$ is the kernel size), and can be trained to recognise various patterns depending on the task at hand. A learnable scalar offset, or bias term, is applied to the output of each matched filter. 

Both modules employ three one-dimensional convolutional layers, which are applied sequentially. Inbetween each convolutional layer, a nonlinear activation function is applied to each sample of each channel. The activation function is chosen to be the Recified Linear Unit, defined as $\mathrm{ReLU}(x) = \max(0,x)$. Each convolutional layer has a kernel size of three temporal samples, and each layer implements 16 matched filters. Therefore, the projected representation of each temporal segment of the stimulus or the EEG is a 16-channel time-series. The first convolutional layer of the EEG module is implemented as a separable convolution, which means that the weight matrices of the matched filters are constrained to the space of rank-1 matrices. This reduces the number of independent parameters required to implement 16 multi-channel matched filters operating on a 64-channel time-series. Finally, the convolutional layers employ dilated convolutions. In a dilated convolution, the kernel elements are not adjacent to one another, and there are gaps in between them - this increases the receptive field of each matched filter without increasing the number of required parameters. The dilation rate is a hyperparameter which controls the spacing of the kernel elements. For both modules, the first convolutional layer has a dilation rate of 1 (no dilation/adjacent kernel elements), the second has a dilation rate of 3 (there are two temporal samples between kernel elements), and the third has a dilation rate of 9 (there are eight temporal samples between kernel elements). The concept of dilated convolutions is illustrated in Figure~\ref{fig:architecture}b. These hyperparameters are the same as those originally suggested by Accou~\textit{et al.}, and no further hyperparameter tuning was performed~\cite{Accou2021EUSIPCO}. From the hyperparameters, the width of the receptive fields of the convolutional modules can be determined to be 27 temporal samples. Since the envelope-based and speech-FFR-based decoders operate at different sampling frequencies (\qty{64}{Hz} and \qty{512}{Hz} respectively), a width of 27 samples corresponds to timescales of \qty{422}{ms} for the envelope-based decoder, and \qty{53}{ms} for the speech-FFR-based decoder.

In the match-mismatch task, one EEG segment and two stimulus segments (one matched and one mismatched) are presented to the DNN. The two stimulus segments are presented as an ordered pair, and the DNN is trained to recognise the order of the matched segment and the mismatched segment within this ordered pair. The match-mismatch task is therefore a binary classification problem, with the first class containing all examples for which the matched segment precedes the mismatched segment, and the second class containing all examples for which the mismatched segment is positioned first in the ordered pair. The similarity between the projected EEG segment and each projected stimulus segment is assessed through a cosine similarity operation. This is implemented by first normalising each channel of each representation to have unit magnitude, and then matrix-multiplying the two 16-channel time-series together (dot product of normalised vectors). This procedure results in an ordered pair of $16\times16$ matrices of cosine similarity scores.

The DNN is required to make a binary classification decision based on this ordered pair of matrices. This is achieved by performing an element-wise subtraction of the second matrix from the first. The resulting matrix elements are then flattened and fed into a single output neuron with no bias term. In other words, a scalar linear combination of those matrix elements is formed. To obtain the final output of the DNN, a sigmoid function is applied to the resulting scalar. The sigmoid function produces a number between 0 and 1 which is taken to represent the probability that the inputs to the DNN belong to the first class (the matched stimulus segment precedes the mismatched stimulus segment). These operations are represented graphically in Figure~\ref{fig:architecture}c.

Our architecture is a modified version of that employed by Accou~\textit{et al}. and Puffay~\textit{et al}.~\cite{Accou2021, puffay22_interspeech}. Whereas we used a separable convolutional layer at the start of the EEG module to handle the large number of input EEG channels, those authors first took several channel-wise linear combinations of the EEG signals (spatial filtering) and then proceeded to use three layers of ordinary, non-separable dilated convolutions. Also, rather than taking an element-wise difference of the two similarity matrices, those authors fed the (flattened) elements of both matrices to the output neuron (which had a bias term). An advantage of our approach is that the predicted class probabilities become exactly symmetric with respect to a change in the ordering of the matched and mismatched stimulus segments. The architecture of Accou~\textit{et al}. and Puffay~\textit{et al} must instead learn this symmetry through training.

\subsection{Training procedure}

The performance of a binary classifier can be assessed through the binary cross-entropy (BCE) loss function, defined as:

\begin{equation}
\centering
L = -[y \log (\hat{y}) + (1-y)\log (1-\hat{y})],
\end{equation}

where $\hat{y}$ is the predicted probability that an input example belongs to the first of two classes, and $y$ is the true target label. Classifiers which make confident and accurate predictions achieve low BCE scores. Since the BCE loss function is differentiable (unlike the classification accuracy metric), it can be minimised with respect to the parameters of the classifier using gradient-based optimizers. In fact, it can be shown that through minimising the binary cross-entropy loss function, the likelihood function of the class probabilities is maximised with respect to the classifier's parameters~\cite{BishopPRML}.

We trained the decoders by minimising the BCE loss function with respect to the decoder parameters. Minimisation of the loss function was performed using the Adam optimizer, which is based on the stochastic gradient descent algorithm~\cite{KingmanAdam}. In stochastic gradient descent, small batches of training examples are fed through the decoder, and the outputs are used to estimate the gradient of the BCE loss function with respect to the parameters of the decoder. Then, a small weight update is added to each parameter: the weight update is proportional to the magnitude of the estimated gradient, but it has the opposite sign. The scale of the weight update is controlled by a multiplicative hyperparameter called the learning rate. When small batch sizes are used, the gradient estimates employed in SGD are noisy and the rate of convergence is slow. The Adam optimizer tackles this by employing an exponentially-weighted moving average of the estimated gradients. Furthermore, for each parameter, it scales the learning rate adaptively by a factor which is inversely related to the variance (noisiness) of the BCE gradient estimate for that parameter.

Batches of 128 training examples (with balanced target classes) were presented sequentially to the decoder during training. A training epoch is said to have occurred once all possible examples have been presented to the decoder exactly once. Stochastic gradient-based optimizers do not converge that quickly, so training is usually required to continue for many epochs before a low and stable value of the loss function is achieved. In our experiments with the development dataset, we found that it was beneficial to employ learning rate scheduling: every seven epochs, the learning rate hyperparameter was reduced by a factor of ten. We also used an early stopping protocol with a patience of five epochs: after each training epoch, we evaluated the average BCE loss on the validation portion of the development dataset. If the validation loss had not decreased within five epochs, training was terminated and the decoder parameters which achieved the best validation loss were saved.

Usually, EEG decoding accuracies are higher when information can be integrated across long temporal segments of EEG signals~\cite{monesi2020ICASSP, Geirnaert2020TransNSRE, Accou2021}. For the Auditory EEG Decoding Signal Processing Grand Challenge, the organisers selected a segment length of three seconds in duration: this is long enough for significant decoding accuracies to be achieved, but short enough to avoid ceiling effects due to teams achieving near-perfect classification accuracies. In this work, we also used segments of \qty{3}{s} in duration.

Puffay~\textit{et al.} recommend taking care when choosing which mismatched segments to use during training: ideally, matched and mismatched segments should be drawn from the same distribution (i.e. the mismatched segments should be `hard negatives')~\cite{puffay2023relating}. This is intended to promote proper learning of the association between the speech features and the EEG signals, rather than learning of the differences between the characteristics of the matched and mismatched speech segments. In particular, Puffay~\textit{et al.} suggest selecting each mismatched segment from the same speech material as the corresponding matched segment, but at a fixed delay relative to the matched segment: a spacing of one second between the end of the matched segment and the onset of the mismatched segment was demonstrated to work well, and we adopt this scheme for both training and evaluating our decoders.

There was an overlap of \qty{2}{s} between consecutive training examples. One effect of this is that every matched segment in the training dataset also appears as a mismatched segment. Puffay~\textit{et al.} noted that this can help to mitigate overfitting, since the decoder can no longer simply learn to associate specific pairs of stimulus segments with their associated labels~\cite{puffay2023relating}.

\subsection{Overview of experiments}

We trained two types of decoder to solve the match-mismatch task. The first type (envelope-based decoders) related the EEG recordings to the slow amplitude fluctuations in the speech streams. The second type (FFR-based decoders) related the EEG recordings to high-frequency periodicity in the speech envelopes. Various experiments were performed using these two types of decoders.

\subsubsection{Model averaging and comparison to baseline}

There are two main sources of randomness in the training procedure. Firstly, the initial parameterisation of the decoder was random: the weights were drawn from Glorot uniform distributions and the bias terms were initialised to zero~\cite{pmlr-v9-glorot10a}. Secondly, the order in which training examples were presented to the decoders was random. Although training examples from a particular EEG trial were presented in order, the order in which trials were selected was random and shuffled after each epoch. We investigated the impact of these sources of randomness by training 100 distinct instances of both the envelope-based decoder as well as the FFR-based decoder. We also studied the benefits of averaging sigmoid outputs taken from multiple trained decoder instances of the same type. In our subsequent analyses, we always use the average of the sigmoid outputs of all 100 trained decoder instances. We refer to the decoders that use averaged sigmoid outputs as `averaged decoders'. Please note that the deep learning framework which we employed (PyTorch version 2.0.1) is also affected by other sources of randomness search as non-deterministic algorithms, and these cannot be controlled by setting the random seed alone~\cite{NEURIPS2019_9015}.

The envelope-based decoder employed in this work is a modified version of the decoder proposed by Accou~\textit{et al.}~\cite{Accou2021EUSIPCO}, as described in Section~\ref{sec:methods-dnns} of the Materials~and~Methods section. Therefore, it is important to establish  how our modifications to the baseline architecture of Accou~\textit{et al.} impacts the match-mismatch decoding performance. To this end, we used the same training procedure to train 100 instances of the population baseline decoder (i.e. without fine-tuning), which were compared against the 100 instances of our envelope-based decoder. Firstly, we performed statistical tests to compare the performances of the two groups of 100 instances using the development dataset. Secondly, we evaluated the performance of the 100-instance averaged baseline decoder using each of the datasets considered in this work, and report the results alongside those obtained with our proposed decoders in Table~\ref{table:heldout}.

\subsubsection{Effect of speaker pitch}

The speech-FFR is known to be modulated by speaker pitch~\cite{forte2017human, Kulasingham2020}. In particular, the speech-FFR is weaker when elicited by higher-pitched voices. We expected the accuracy of the FFR-based decoders to reflect the pitch of the speech material. To assess this, we estimated the mean pitch of each audiobook or podcast in the testing portion of the development dataset using the Praat software~\cite{Praat}. We also calculated the mean classification accuracy (taken across participants) of the averaged decoders. The accuracies and pitches were compared using Pearson's correlation coefficient. We also compared the decoding accuracies achieved for the male-narrated speech material against those achieved for the female-narrated speech material.

\subsubsection{Decoder fine-tuning}

Electroencephalography signals are highly participant-specific, since they depend on intrinsic factors such as the anatomy of the participant in question, as well as extrinsic factors such as the placement of the electrode cap and the impedances of the skin-electrode interfaces (which usually will be session-dependent). Some prior studies have shown that it can be beneficial to fine-tune a trained instance of the population decoder to individual participants~\cite{Accou2021, monesi2020ICASSP}. Inspired by these studies, for each type of decoder we selected for fine-tuning the instance which achieved the best accuracy when evaluated on the testing portion of the development dataset. Then, for each participant in the development dataset, we resumed the training of the decoder using data from that participant only.
	
\subsubsection{Composite decoder}

We combined our averaged envelope-based decoder with the averaged FFR-based decoder using a linear classifier (linear discriminant analysis, LDA), which operated on the sigmoid outputs of both decoders to predict a final class estimate. The LDA classifier was trained using the testing portion of the development dataset. For participants in the heldout dataset who also appeared in the development dataset, we additionally formed a `fine-tuned' composite decoder. This decoder utilised the same population-based LDA classifier as before (with the same parameters). However, the sigmoid outputs of the fine-tuned decoders were used in place of those of the averaged decoders. Based on our experience gained during the ICASSP 2023 Auditory EEG Decoding SPGC, we decided against using participant-specific LDA classifiers, since these were found to generalise poorly to the heldout dataset.

\subsubsection{Generalisation of the decoders to a distinct dataset}

Finally, we assessed how well both types of averaged population-based decoder, as well as the composite decoder, could generalise to the completely distinct dataset of Etard~\textit{et al.}~\cite{EtardDataset}. This dataset consists of EEG measurements recorded from native English-speaking participants who listened to audiobooks in several different listening conditions: speech in quiet; speech in babble noise; speech in a foreign language; and two-talker competing-speakers conditions. We assessed the match-mismatch classification accuracy of the decoders in all of the listening conditions, for both the target speaker and the ignored speaker in the case of the competing-speakers conditions. We also performed auditory attention decoding with these decoders, by replacing the mismatched segment with the temporally-aligned segment of the ignored speech stream. The matched segments were kept as the temporally-aligned segments of the attended speech stream.

\section{Results}

\subsection{Random seed initialisation and model averaging}

Various sources of randomness have an impact on the final state of a trained decoder. We trained 100 instances of both the envelope-based decoder as well as the FFR-based decoder. Each instance was trained using data from all participants in the development dataset. The random number generators used during decoder training were initialised with a different seed for each decoder instance.

\begin{figure*}[ht!]
    \centering
    \subfigure[]{\includegraphics[width=0.49\linewidth]{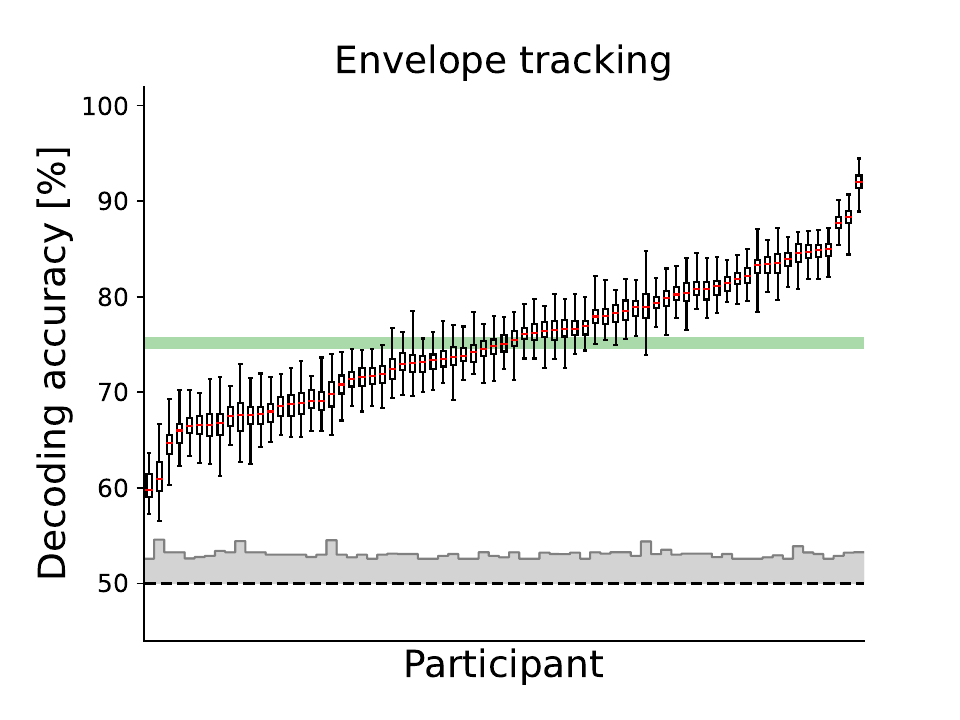}}
    \subfigure[]{\includegraphics[width=0.49\linewidth]{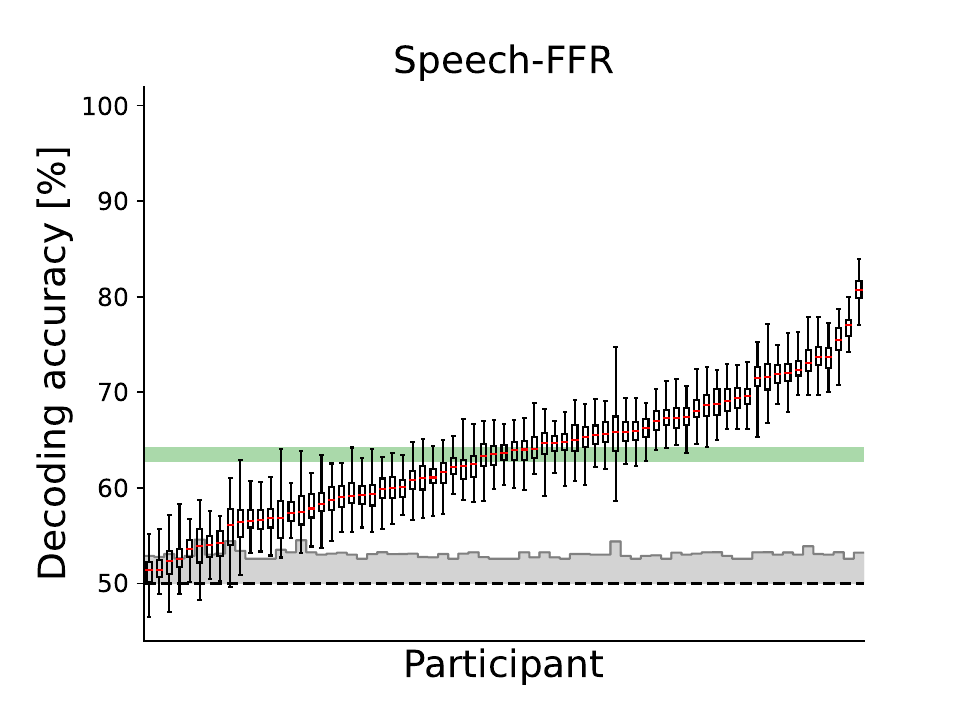}}

    \caption{The decoding accuracies for individual participants in the development dataset. Participants are ordered by median classification accuracy. Each boxplot shows the spread of accuracies achieved by 100 decoder instances, which were trained with different random seeds. The whisker-to-whisker distance represents the range of the data. The grey region shows the upper limit of the \qty{95}{\%} confidence interval of a random binary classifier. Additionally, for each decoder we also calculated the mean accuracy across all participants. The range of the 100 mean classification accuracies is indicated by the region highlighted in green. a) Results for the envelope-based decoders. b) Results for the FFR-based decoders.}
    \label{fig:sub-level-accs}
\end{figure*}

We evaluated the decoders using the testing portion of the development dataset. Different decoder instances achieved a large range of accuracies when evaluated on data from individual participants, as shown in Figure~\ref{fig:sub-level-accs}. The average range of accuracies for individual participants is 6.7 percentage points for the envelope-tracking based decoder, and 7.9 percentage points for the FFR-based decoder. On average, the standard deviations of the decoding accuracies were 1.3 and 1.5 (in percentage points), respectively. For each decoder instance we also calculated an overall mean classification accuracy, by taking the mean of the classification accuracies achieved for all 71 participants in the development dataset. Remarkably, compared to the accuracies achieved for individual participants, the participant-average classification accuracy was much more narrowly distributed. For the envelope-based decoder, the range of the 100 participant-average classification accuracies was \qtyrange{74.6}{75.8}{\percent}; for the FFR-based decoder, this range was \qtyrange{62.7}{64.3}{\percent}.

\begin{figure*}[ht!]
    \centering
    \footnotesize	
    \subfigure[]{\includegraphics[width=0.49\linewidth]{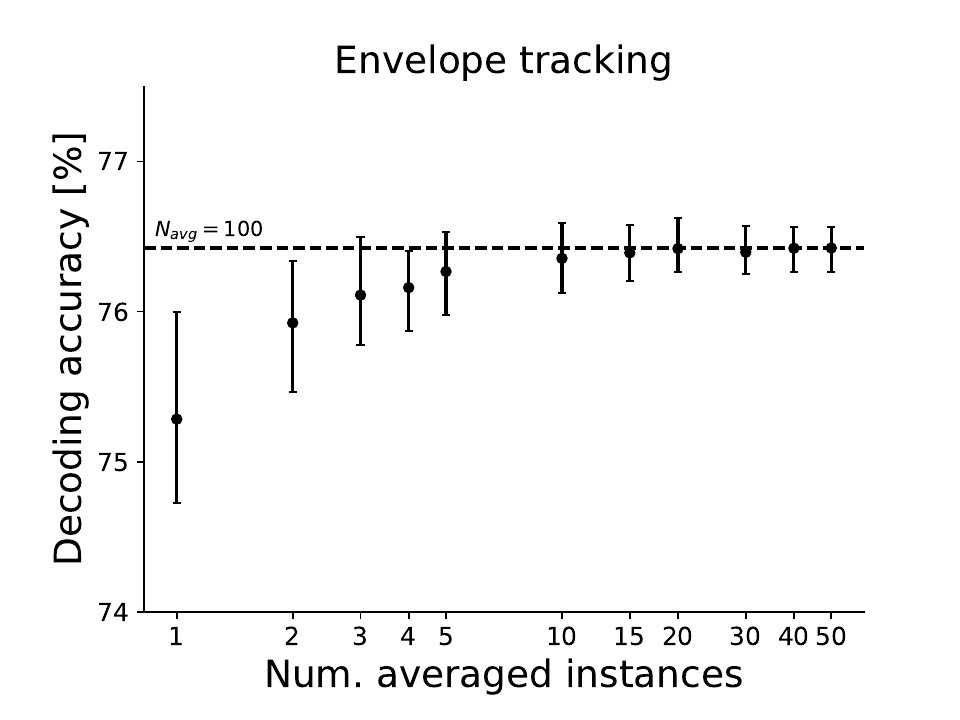}}
    \subfigure[]{\includegraphics[width=0.49\linewidth]{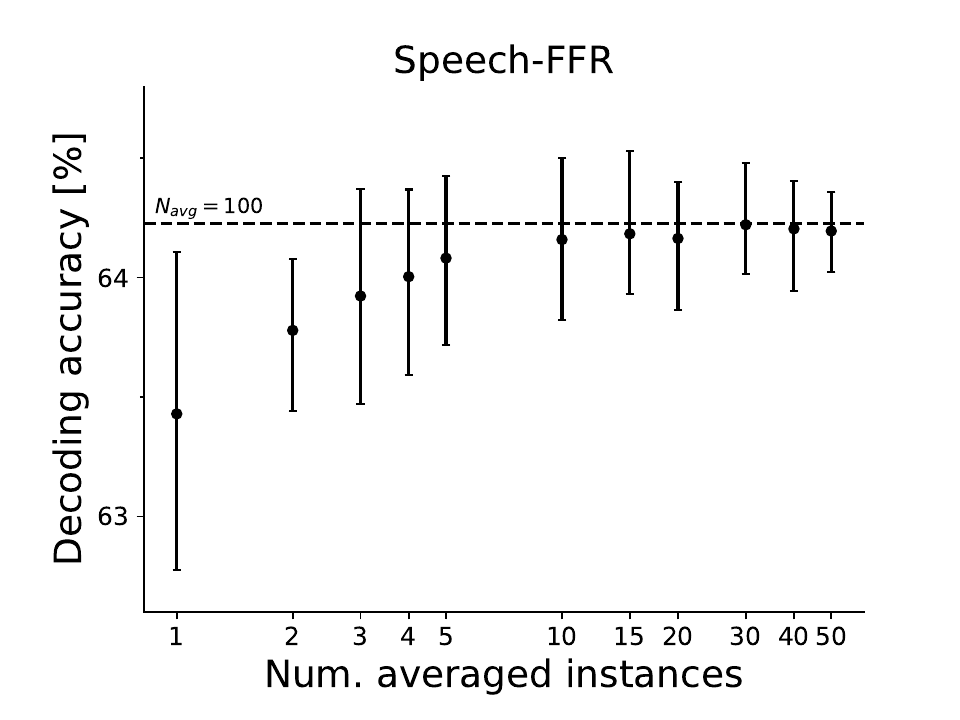}}

    \caption{Averaging of decoder sigmoid outputs. The bootstrapped mean and range of the participant-average decoding accuracy (as evaluated on the testing portion of the heldout dataset) is shown against the number of decoders used in the average. a) Effect of averaging the sigmoid outputs of the envelope-based decoder. b) Effect of averaging the sigmoid outputs of the FFR-based decoder. In both cases, the mean decoding accuracy increases with the number of averaged decoder instances; a plateau is achieved by around ten averaged instances. The dotted lines represent the participant-average decoding accuracy achieved by the 100-instance-average decoders.}
    \label{fig:decoder-averagings}
\end{figure*}

A simple ensembling procedure was used to take advantage of the apparent diversity within the two sets of 100 trained decoder instances. Specifically, we selected a number of decoder instances of a particular type at random and without replacement, and averaged their sigmoid outputs (predicted class probabilities) for a given input example. By doing so, we formed ensembled decoders which were more accurate than any of the constituent decoders. The effect of averaging different numbers of decoders was assessed using a bootstrapping procedure; the number of decoder instances ($n$) to be averaged was varied, and for each $n$ that was considered we drew 50 sets of $n$ trained decoder instances. Figure~\ref{fig:decoder-averagings} shows the mean of the classification accuracy, as well as its range, against the number of averaged instances. By averaging ten instances of the decoders, the participant-average decoding accuracy could be improved from \qty{75.3}{\percent} to \qty{76.4}{\percent} (1.3 percentage points) for the envelope-tracking based decoder, and it was improved from \qty{63.4}{\percent} to \qty{64.2}{\percent} (0.8 percentage points) for the FFR-based decoder. On average, the performance of the decoders does not increase when even more decoder instances are combined in this way.

\subsection{Effect of speaker pitch}
\label{sec:pitch}

\begin{figure*}[ht!]
    \centering
    \footnotesize	
    \subfigure[]{\includegraphics[width=0.49\linewidth]{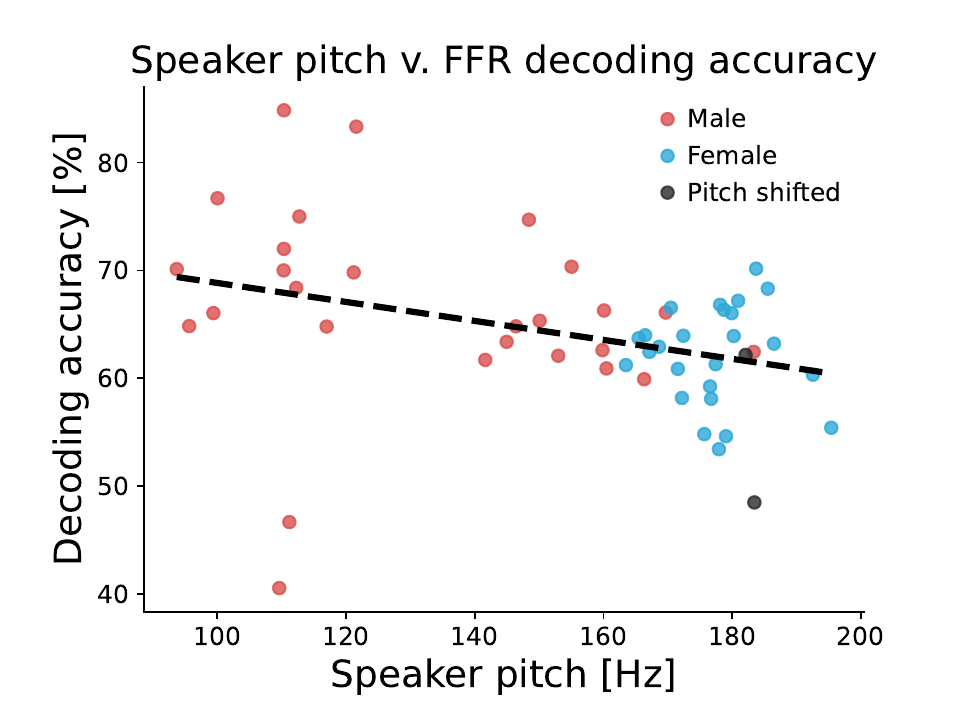}}
    \subfigure[]{\includegraphics[width=0.49\linewidth]{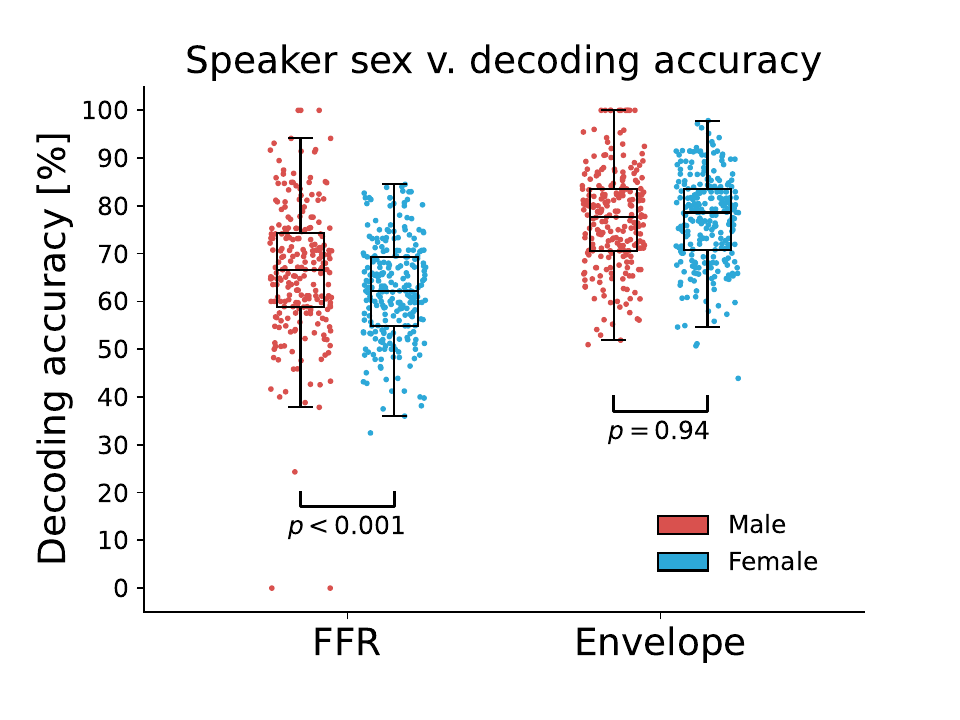}}

    \caption{The accuracy of the averaged FFR-based decoder varies with speaker pitch and sex. a) The average accuracy across participants who listened to a particular audiobook or podcast is plotted against the mean pitch of that audiobook or podcast. A regression line is also shown. Two of the male-narrated audiobooks were synthetically pitch-shifted into the range of female-narrated speech, and the corresponding scatter points are shaded grey. b) Each datapoint represents the classification accuracy of the FFR-based decoder or the envelope-based decoder, as evaluated on the testing portion of each trial in the development dataset. For the FFR-based decoder, there was a statistically significant difference between the classification accuracies for male-narrated and female-narrated speech material.}
    \label{fig:pitch-analysis}
\end{figure*}

In order to assess the relationship between speaker pitch and decoding accuracy, the mean pitch of each audiobook or podcast in the development dataset was computed. Then, using the testing portion of the development dataset, the classification accuracy for each participant who listened to that audiobook or podcast was determined. The average amongst those decoding accuracies was then calculated. The set of average accuracies was subsequently correlated against the set of mean speaker pitches (Pearson correlation coefficient). For the speech-FFR decoder, there was a statistically significant correlation of $R=-0.34$ between the two variables ($p=0.01$, exact single-tailed test using all $N=57$ stories assuming both variates are jointly normal). We also computed a least-squares linear fit between the two variables, which had an intercept of 77.63 and a slope of \qty{-0.088}{\per\hertz}, as shown in Figure~\ref{fig:pitch-analysis}. For the envelope-based decoder, there was no statistically significant correlation ($R=-0.18$, with $p=0.18$.)

Male-narrated speech typically has a lower pitch than female-narrated speech. As a second assessment of the relationship between speaker pitch and classification accuracy, each narrative was labeled as either male-narrated or female-narrated. Then, for each participant who listened to that narrative, a classification accuracy was computed (using the testing portion of the development dataset). The two groups of classification accuracies (for male- and female-narrated speech, respectively) were compared via a two-tailed, unpaired t-test. For the FFR-based decoder, there was a highly significant statistical difference between the two groups ($p=0.0003$), whereas there was no significant difference between the two groups for the envelope-based decoder ($p=0.94$).

\subsection{Decoder fine-tuning}

\begin{figure}
    \centering
    \footnotesize	
    \includegraphics[width=\linewidth]{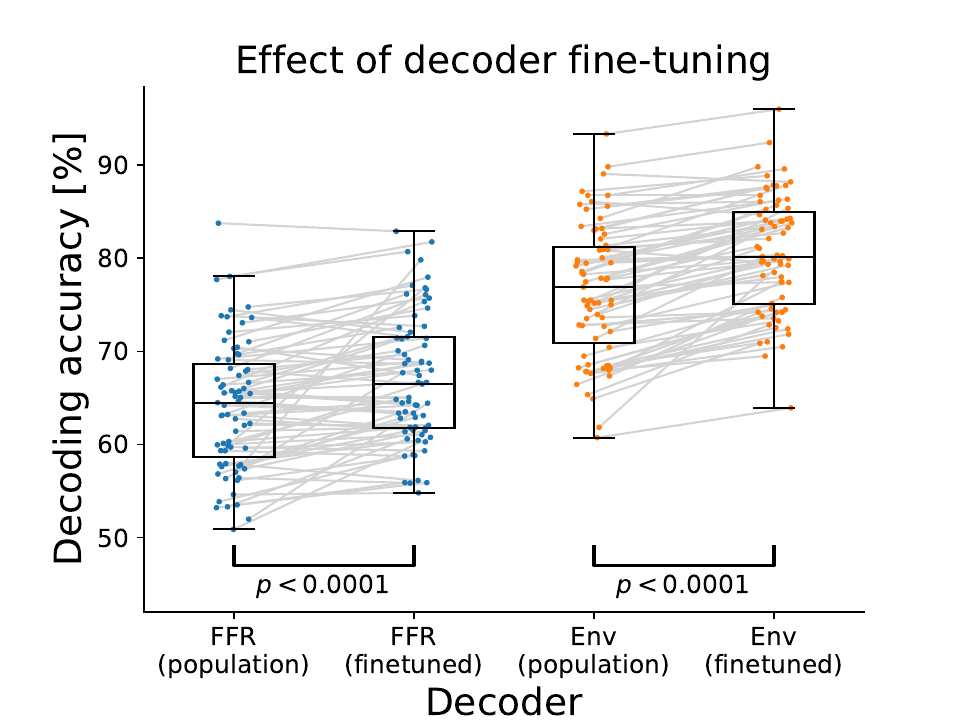}

    \caption{Comparison between the averaged decoders, which were trained on all 71 participants in the development dataset, and the decoders which were fine-tuned to each of those participants. Each datapoint shows the classification accuracy for a particular participant and decoder, as evaluated using the testing portion of the development dataset. Grey lines connect the accuracies achieved for individual subjects.}
    \label{fig:fine-tuning}
\end{figure}

One fine-tuned decoder was produced for each of the 71 participants who featured in the development dataset. Fine-tuning was performed by taking the best population decoder from the set of 100 trained instances (as assessed by its classification accuracy on the testing portion of the development dataset), and resuming the training of this decoder using data from just one participant. The performance of the fine-tuned decoders was compared against the performance of the averaged population decoders using the testing portion of the development dataset, and the results are shown in Figure~\ref{fig:fine-tuning}. Fine-tuning offered a highly statistically significant improvement in decoding accuracy for both the speech-FFR decoder as well as the envelope-based decoder, when compared against the respective averaged decoders ($P<<0.0001$, single-tailed paired t-tests). The significance of the improvement in decoding accuracy was replicated when the decoders were evaluated using the heldout dataset (Figure~\ref{fig:heldout-eval}). To quantify the effect size of the fine-tuning, we report the \qty{95}{\percent} confidence interval on the mean of the paired differences for each type of decoder. For the envelope-based decoder, this was the interval \qtyrange{0.0}{7.9}{\percent}, and for the speech-FFR decoder this was the interval \qtyrange{0.0}{5.8}{\percent}.

\subsection{Composite decoder}
\label{sec:composite}

\begin{figure*}[ht!]
    \centering
    \footnotesize	
    \subfigure[]{\includegraphics[width=0.49\linewidth]{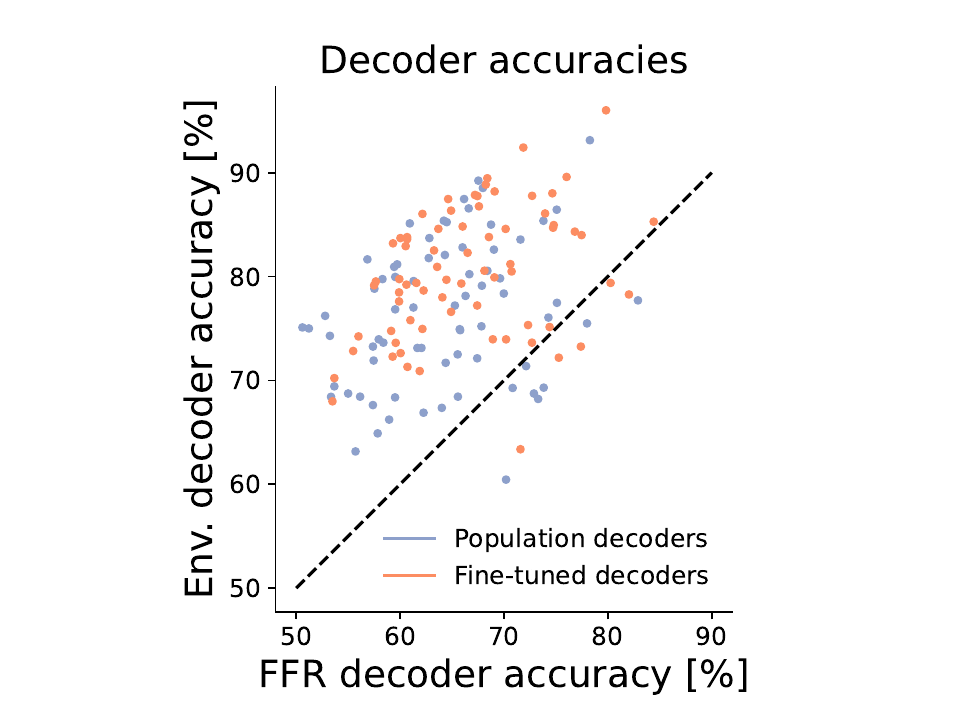}}
    \subfigure[]{\includegraphics[width=0.49\linewidth]{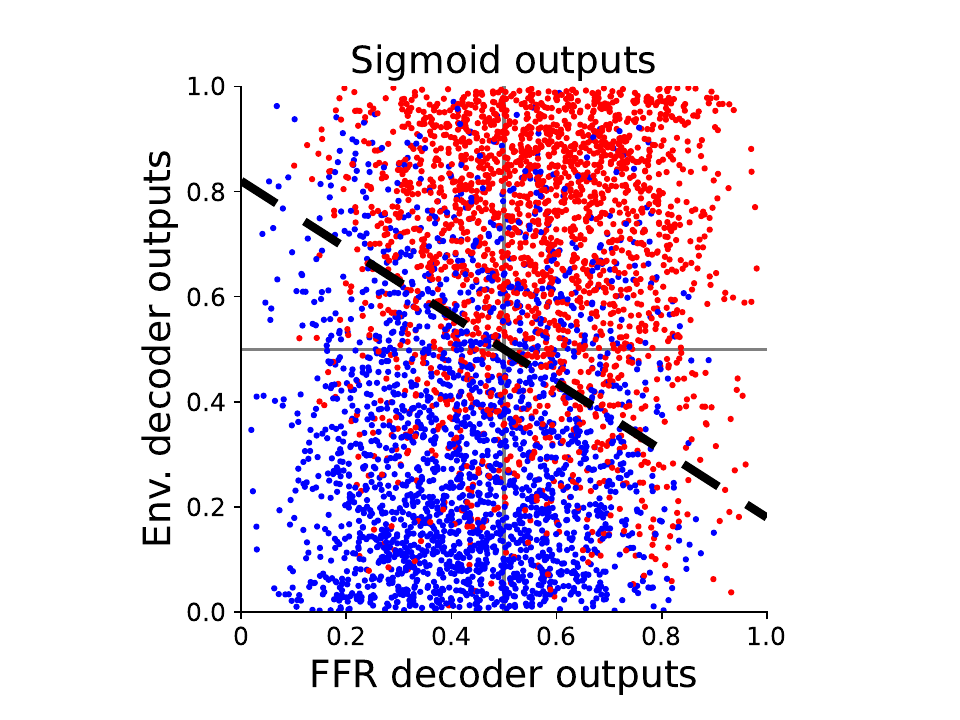}}
    \caption{Comparison between the decoding accuracies and sigmoid outputs of the two types of decoders. a) Decoding accuracies for individual participants, as evaluated on the testing portion of the development dataset. The envelope-based decoders outperform the FFR-based decoders for almost all participants. b) Sigmoid outputs of the two averaged decoders, calculated using the same dataset. A linear classifer (LDA) was trained to predict the true class labels, indicated by the colour of the datapoints, using these sigmoid outputs. The decision boundary of the classifier is designated by the black dashed line.}
    \label{fig:decoder-comparison}
\end{figure*}

For almost all participants, the envelope-based decoders achieved higher classification accuracies than the FFR-based decoders when evaluated using the testing portion of the development dataset (see Figure~\ref{fig:decoder-comparison}a). The classification accuracies of the two averaged decoders were not that correlated ($R=0.286$, Pearson's correlation coefficient), and nor were their sigmoid outputs ($R=0.233$). We decided to combine the two averaged decoders via a linear classifier (LDA), which was trained on the testing portion of the development dataset to predict the true class label from the sigmoid outputs of both decoders. As shown in Figure~\ref{fig:decoder-comparison}b, the decision boundary of the linear classifier is defined by the contour $0.39p_f + 0.61p_e = 0.5$, where $p_f$ and $p_e$ are the sigmoid outputs of the averaged FFR-based and envelope-based decoders respectively. Therefore, the linear classifier assigns weights of \qty{39}{\percent} and \qty{61}{\percent} to these respective decoders. Figure~\ref{fig:heldout-eval}b shows that the composite decoder provides a reliable improvement over the averaged envelope-based decoder when evaluated on the heldout dataset.

A fine-tuned composite decoder was formed by using the same linear classifier (retaining its parameters), but using as inputs the sigmoid outputs of the fine-tuned decoders rather than the averaged decoders. Figure~\ref{fig:heldout-eval} shows that the population and fine-tuned composite decoders achieved the highest classification accuracies (for seen and unseen participants respectively) of any decoders considered in this work. This result is also summarised in Table~\ref{table:heldout}.

\subsection{Evaluation of all decoders on the heldout dataset}

\begin{figure*}
    \centering
    \footnotesize	
    \subfigure[]{\includegraphics[width=.49\linewidth]{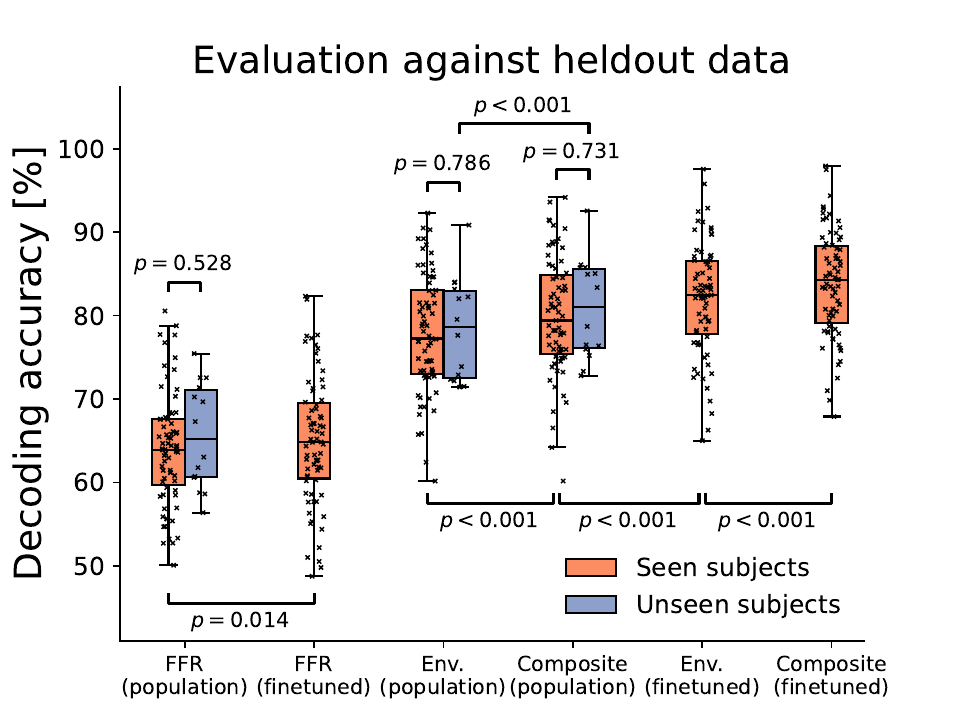}}
    \subfigure[]{\includegraphics[width=.49\linewidth]{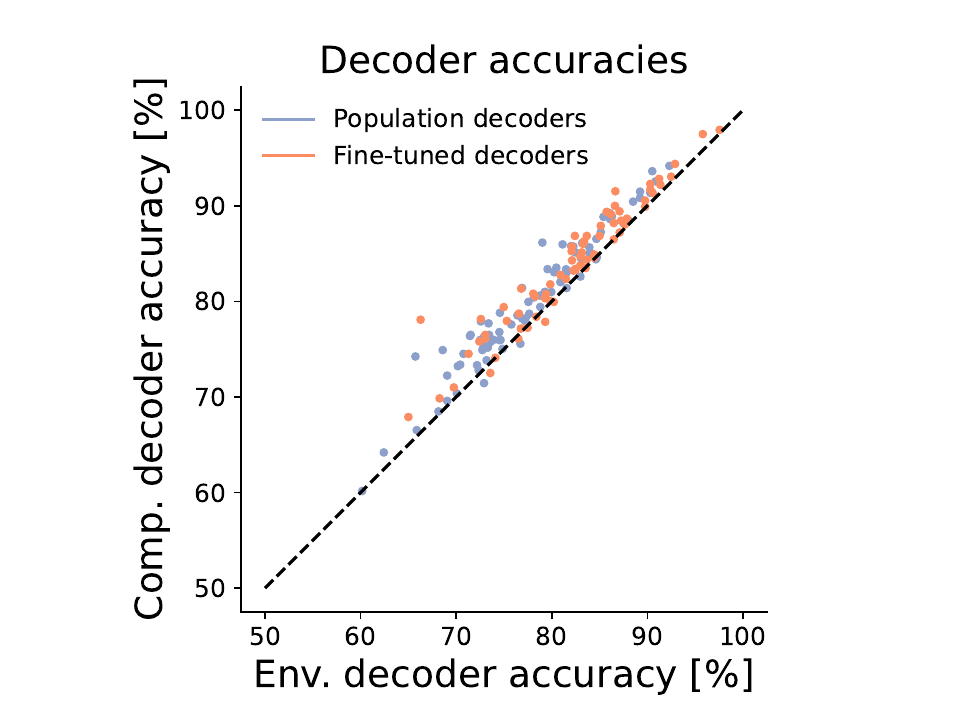}}
    \caption{Evaluation of the decoders using the heldout dataset. (a) Comparison between the performances of all of the considered decoders. Datapoints represent the classification accuracy for individual participants. The fine-tuned decoders could only be applied to participants who had already been seen in the development dataset, and hence there are no boxplots corresponding to the application of fine-tuned decoders to unseen participants. All reported p-values were calculated through t-tests and have been FDR-corrected for multiple comparisons. For comparisons between the seen participants and the unseen participants, two-tailed unpaired t-tests were used. Otherwise, all of the tests were single-tailed paired t-tests. (b) Comparison between the composite decoder and the envelope-based decoder, for individual participants. For the population decoders, accuracies for seen and unseen participants are grouped together and shown in blue. The orange datapoints represent accuracies for the finetuned decoders, which could only be evaluated for seen participants.}
    \label{fig:heldout-eval}
\end{figure*}

We evaluated all of the decoders using the heldout dataset, which was completely unseen during the development of the decoders (aside from the results of our four submissions to the Auditory EEG SPGC~\cite{ICASSPTHORNTON, jalilpour2022icassp}). This dataset consisted of data from participants who had already been seen in the development dataset, as well as data from completely unseen participants. The results are presented in Figure~\ref{fig:heldout-eval}, and summarised in Table~\ref{table:heldout}. Overall, the population decoders generalised to unseen participants extremely well. Composite decoders performed better than their constituent decoders, and fine-tuned decoders performed better than averaged population decoders. The fine-tuned composite decoder achieved a particularly high mean accuracy of 83.79\%, calculated across seen participants.

\begin{table*}
\centering
   \begin{tabular}{ |l||c|c|c||c|c|c| }
 \hline
 & \multicolumn{3}{c||}{SparrKULee} & \multicolumn{3}{c|}{ICL dataset} \\
 \hline
 Decoder& Development & Heldout (seen) & Heldout (unseen)& Speech-in-quiet & Foreign language & Attention decoding\\
 \hline
 FFR (population)  & $64.22\pm 1.65$ & $63.98\pm 1.58$ & $65.64\pm 3.52$ & $60.94\pm 2.19$ & $61.49\pm 4.61$ & $52.08\pm 1.36$\\
 FFR (fine-tuned)  & $66.65\pm 1.67$ & $65.01\pm 1.79$ & -               & -               & - & -\\
 Env. (population) & $76.42\pm 1.64$ & $77.86\pm 1.61$ & $78.41\pm 3.42$ & $81.27\pm 2.99$ & $79.13\pm 3.12$ & $62.86\pm 2.25$ \\
 Comp. (population)& -               & $79.95\pm 1.65$ & $80.89\pm 3.36$ & $81.86\pm 3.01$ & $80.39\pm3.04$ & $62.60\pm 2.45$ \\
 Env. (fine-tuned) & $80.39\pm 1.47$ & $81.98\pm 1.60$ & -               & -               & - & -\\
 Comp. (fine-tuned)& -               & $83.79\pm 1.51$ & -               & -               & - & -\\
 \hline
 Accou~\textit{et al.} (population)  & $76.15\pm 1.64$ & $77.47\pm 1.64$ & $78.02\pm 3.02$ & $80.01\pm 3.04$ & $78.57\pm 3.40$   & $62.75\pm2.23$\\
 \hline
\end{tabular}
\vspace*{5mm}
 \caption{Comparison between all decoders using the various datasets considered in this work. The 95\% confidence intervals on the participant-average decoding accuracies (mean $\pm$ 95\% margin of error) are reported. All of the population decoders were averaged decoders (i.e. the predicted class probabilities are averaged across 100 trained decoder instances). The fine-tuned decoders were individualised for participants who featured in the development dataset, and therefore could not be applied to the unseen participants in the heldout dataset. The composite decoders employed a linear classifier to combine the sigmoid outputs of the envelope-based and FFR-based decoders. The linear classifier was trained using the testing portion of the development dataset, and hence the composite decoders could not be evaluated on this dataset. The decoders were applied to the completely distinct ICL dataset, which includes EEG recordings acquired under various listening conditions. Results shown for this dataset include the average match-mismatch classification accuracy for the speech-in-quiet condition, the foreign-language speech condition, and the average attention decoding accuracy for the competing-speakers conditions.}
  \label{table:heldout}
\end{table*}

\subsection{Generalisation to other datasets}

As shown in Section~\ref{sec:composite}, the population decoders generalised remarkably well to participants who were not represented in the development dataset. We assessed the generalisation capabilities of the population decoders further by evaluating them to the ICL dataset, which was completely unseen during the training and development of the decoders~\cite{EtardDataset}. The EEG electrodes were placed slightly differently in this dataset as compared with SparrKULee. The ICL dataset also contains EEG recorded under several listening conditions: speech in quiet, speech in noise, foreign-language speech, and competing-speakers. For the speech-in-noise and speech-in-quiet conditions, audiobooks were narrated by a female speaker with a mean pitch of \qty{182}{Hz}. The foreign-language speech material was narrated by another female speaker with the same mean pitch.

\begin{figure*}[ht!]
    \centering
    \subfigure[]{\includegraphics[width=0.49\linewidth]{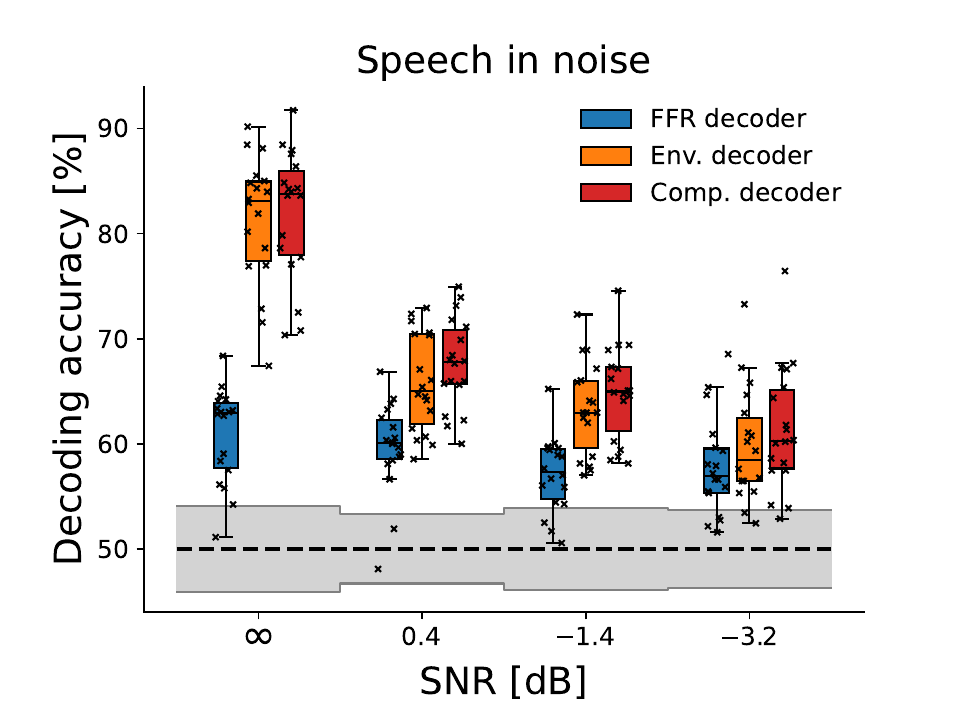}}
    \subfigure[]{\includegraphics[width=0.49\linewidth]{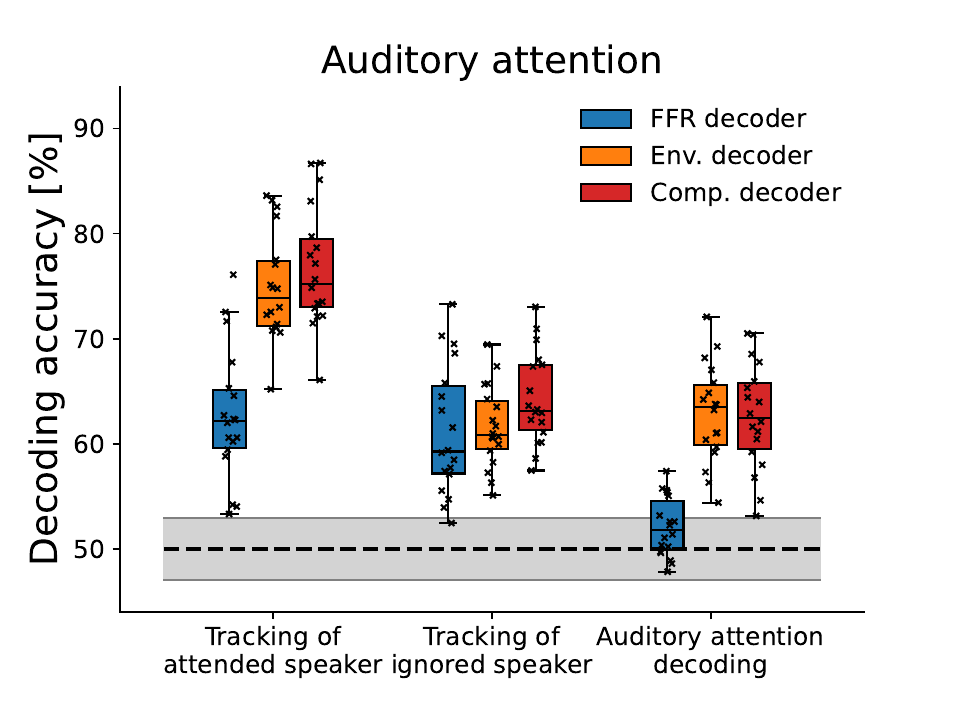}}

    \caption{Performance of the averaged decoders when applied to the ICL dataset. The grey region indicates the 95\% confidence interval of a random classifer. a) Evaluation on speech-in-quiet and speech-in-noise data. Each datapoint represents the decoding accuracy for an individual participant. The envelope-based decoder generalises extremely well to the speech-in-quiet data. b) Evaluation on competing-speakers data. The match-mismatch classification accuracies of the decoders are reported in the first two pairs of boxes - in the first, stimulus segments are drawn from the attended speaker; in the second, they are drawn from the ignored speaker. In the third group, the trained decoders were used as auditory attention decoders.}
    \label{fig:generalisation}
\end{figure*}

For our first case study, the decoders were evaluated on the speech-in-quiet and the speech-in-noise data. The results are shown in Figure~\ref{fig:generalisation}a. The envelope-based decoder generalised remarkably well to the speech-in-quiet data, achieving a mean classification accuracy of \qty{81.27}{\percent}. The statistical difference between this decoding accuracy and the mean decoding accuracy taken over unseen participants in the heldout dataset was borderline significant ($p=0.05$, single-tailed unpaired t-test). Similarly, the mean accuracy of the FFR-based decoder was similar to the accuracy that would be expected for a speaker with a mean pitch of \qty{182}{Hz} based on the least-squares fit reported in Section~\ref{sec:pitch} (expected accuracy using linear fit: 61.6\%; actual 95\% CI on the mean: [58.75\%, 63.13\%]). The mean accuracy of the composite decoder was 81.86\%, which was not statistically greater than that of the averaged envelope-based decoder ($p=0.07$, single-tailed paired t-test.)

The classification accuracy of the envelope-based decoder was considerably degraded when background babble noise was played during presentation of the speech material. Moreover, the mean accuracy of the envelope-based decoder decreased with a decrease in SNR (comparison between high-SNR and medium-SNR conditions: $p=0.0084$; comparison between medium-SNR and low-SNR conditions: $p=0.0006$, single-tailed paired t-tests). The performance of the averaged FFR-based decoder was more robust against the SNR of the speech material (comparison between speech-in-quiet and high-SNR conditions: $p=0.1663$; comparison between high-SNR and medium-SNR conditions: $p=0.0183$; comparison between medium-SNR and low-SNR conditions: $p=0.8131$.)

For our second case study, we applied the same decoders to the competing-speakers data. The match-mismatch classification accuracy of the decoders was assessed, first by taking both stimulus segments from the attended speech stream, and then by using the ignored speech stream instead. There was a stark decrease in the classification accuracies of both the averaged envelope-based decoder as well as the composite decoder when the ignored speech stream was used instead of the attended speech stream. As shown in Figure~\ref{fig:generalisation}b, the averaged FFR-based decoder achieved similar match-mismatch decoding accuracies for both the attended speaker as well as the unattended speaker. In fact, for this decoder there was no statistical difference between the decoding accuracies for the two speakers ($p=0.1081$, two-tailed paired t-test). We also performed an auditory attention decoding experiment. Temporally-aligned segments of the attended speech stream served as the `matched' segments. For the `mismatched segments', temporally-aligned segments of the ignored speech stream were used. All three auditory attention decoders achieved attention decoding accuracies which were significantly greater than 50\% in this experiment (95\% confidence intervals on the mean attention decoding accuracy: averaged FFR-based decoder - [50.72\%,53.44\%]; averaged envelope-based decoder - [60.61\%,65.11\%]; composite decoder - [60.16\%, 65.05\%]). Evidently, the composite decoder did not outperform the envelope-based decoder in the auditory attention decoding experiment.

We also evaluated the trained decoders using EEG collected in the foreign-language speech listening condition, in which the speech material was presented in quiet. As in the English conditions, the speaker had a mean pitch of \qty{182}{Hz}. The population decoders generalised very well to this data, with the averaged FFR-based decoder achieving a mean accuracy of 61.49\%, the averaged envelope-based decoder achieving 79.13\%, and the composite decoder achieving 80.39\%.

We performed two types of statistical tests to compare these decoding accuracies to those achieved in the English speech-in-quiet listening condition. Firstly, we used single-tailed, unpaired t-tests to compare the two groups of decoding accuracies. Next, we considered only the participants in the ICL dataset who took part in both the English and Dutch EEG sessions, and performed paired single-tailed t-tests to compare the two groups of decoding accuracies. We performed these tests for each of the averaged FFR-based decoder, the envelope-based decoder, and the composite decoder. None of the tests returned a positive results (all $p$ values were larger than $0.1$.)

Finally, we used the ICL dataset to assess how the decoding accuracies are impacted by the length of the EEG and speech-feature segments. So far, only results for segments of \qty{3}{s} in duration have been reported in this work. It is well-established, however, that by using longer segment lengths, higher decoding accuracies may be achieved~\cite{Accou2021, geirnaert2020neurosteered}. The decoders which were trained using a segment duration of \qty{3}{s} using the development dataset can be evaluated using longer segment lengths - here, we considered lengths corresponding to durations of \qty{5}{s} and \qty{10}{s}.  The results for the English speech-in-quiet condition, as well as the auditory attention decoding conditions, are shown in Table~\ref{table:seglen}. In line with previous studies, the participant-average decoding accuracies increase reliably  with increasing segment duration, except for those of the averaged speech-FFR decoder when applied to the auditory attention decoding task (one-way repeated measures ANOVA: $p=0.3019$.)

\begin{table}[h!]
\centering
   \begin{tabular}{ |l||c|c|c| }
 \hline
  & \multicolumn{3}{c|}{Match-mismatch (speech-in-quiet) segment length}\\
  \hline
 Decoder& \qty{3}{s} & \qty{5}{s} & \qty{10}{s}\\
 \hline
 FFR      & $60.94\pm 2.19$ & $63.70\pm 2.59$ &  $68.81\pm 3.45$\\
 Envelope & $81.27\pm 2.99$ & $87.79\pm 3.00$ &  $94.72\pm 2.20$\\
 Composite& $81.86\pm 3.01$ & $88.34\pm 2.72$ &  $95.28\pm 2.00$\\
 \hline

\multicolumn{4}{c}{}\\

\hline

& \multicolumn{3}{c|}{Auditory attention decoding segment length}\\
\hline
 Decoder& \qty{3}{s} & \qty{5}{s} & \qty{10}{s}\\
 \hline
 FFR      & $52.08\pm 1.36$ & $52.58\pm 1.82$ &  $53.02\pm 2.84$\\
 Envelope & $62.86\pm 2.25$ & $66.49\pm 2.95$ &  $72.82\pm 3.95$\\
 Composite& $62.60\pm 2.45$ & $66.06\pm 3.00$  & $72.12\pm 4.07$\\
 \hline
\end{tabular}
\vspace*{4mm}
 \caption{The effect of segment length on decoding accuracy. The decoders, which were trained using a segment duration of \qty{3}{s}, were evaluated on the ICL dataset using segment durations of \qty{3}{s}, \qty{5}{s}, and \qty{10}{s}. Top: decoding accuracies for the speech-in-quiet condition, in which participants listened to speech in their native language. Bottom: accuracies for the auditory attention decoding experiment, in which participants listened to two competing speakers. Increasing the segment length reliably increased the decoding accuracy. The exception is for the averaged FFR-based decoder in the auditory attention decoding task, for which there was no significant difference between the mean decoding accuracies achieved for any of the three segment durations (one-way repeated-measures ANOVA.)}
 \label{table:seglen}
\end{table}

\subsection{Comparison between the baseline decoder and the envelope-based decoder}
The envelope-based decoder used in this work is a modified version of the baseline decoder proposed by Accou~\textit{et al.}: the spatial filter layer of the baseline decoder was subsumed into the first layer of the EEG module in the envelope-based decoder, which implements a separable convolution. Also, the output layer is modified so that the predicted class probabilities of the candidate speech segments swap when the order of the segments is swapped. To evaluate how the changes to the baseline decoder architecture affect the final decoding accuracies, we first compared 100 trained instances of the baseline decoder against 100 trained instances of the envelope-based decoder using the development dataset. The 95\% confidence intervals on the participant-average classification accuracies were [72.95\%, 76.04\%] and [73.66\%, 76.83\%], respectively, meaning that on average the envelope-based decoder outperformed the baseline decoder by a small margin of $0.75$ percentage points when evaluated against the development dataset. Although this margin is small, the improvement for individual participants was highly statistically significant (95\% confidence interval on the paired differences of the average decoding accuracies, where the average was taken over the 100 decoder instances: [$0.65\%$, $0.86\%$].

We also investigated the accuracy of the 100-instance-averaged baseline decoder (Table~\ref{table:heldout}). For individual participants, the averaged envelope-based decoder continued to outperform the averaged baseline decoder for all three subsets of SparrKULee (statistical tests for the development, heldout (seen), and heldout (unseen) datasets, respectively: $p=0.006$, $p<0.001$, $p<0.001$; single-tailed paired t-tests). For the ICL dataset, there was no such statistical difference between the accuracies of the baseline decoder and the envelope-based decoder for neither the speech-in-quiet task, nor the attention decoding task ($p=0.455$ and $p=0.244$, respectively. Single-tailed paired t-tests). There was a statistically significant difference between the performance of the two decoders for the foreign-language speech condition, with the averaged envelope-based decoder outperforming the averaged baseline decoder ($p=0.032$, single-tailed paired t-test.) 

\section{Discussion}

We have described and developed our auditory EEG decoders which were the winners of the match-mismatch sub-task of the ICASSP Auditory EEG Signal Processing Grand Challenge~\cite{jalilpour2022icassp}. Two types of decoders, which leveraged cortical responses to the speech envelope as well as the envelope-related speech-FFR respectively, were developed. Decoders which were trained with different random seeds exhibited considerable diversity, and we capitalised on this by employing a simple ensembling procedure whereby the sigmoid outputs of distinct decoder instances were averaged together. We have also fine-tuned the decoders to individual participants, further improving their match-mismatch classification accuracies. However, the best performance was achieved when the two different types of decoders were combined into a single composite decoder. Finally, we have demonstrated that the decoders can generalise extremely well to entirely distinct datasets, and can even serve as auditory attention decoders in competing-speakers conditions.

\subsection{Differences between trained decoder instances, decoder averaging}

%% Results to comment on
% 1. Large variability in performance for individual subjects
% 2. Small variability in mean over subjects
% -> quality-diversity
% 3. Averaging helped improve performance by significant margin
% 4. SPGC teams were separated by narrow margins, this simple trick helped our decoders outperform others

Sources of randomness in the training procedure were shown to affect the classification accuracies of the trained decoders. We explored this effect by training 100 instances of both the envelope-based decoder as well as the FFR-based decoder. For individual participants, a marked variability in the classification accuracy was observed in the two groups of decoder instances. The participant-average classification accuracy was shown to vary across the various trained decoder instances by a much smaller margin. Clearly, a considerable degree of diversity is exhibited by both groups of decoders, with some instances achieving higher classification accuracies for particular participants at the expense of other participants.

We exploited the diversity between the different trained decoder instances through averaging of their sigmoid outputs. This improved the decoding accuracy of each type of decoder by approximately 1 percentage point (when evaluated on the testing portion of the development dataset). The averaging method was useful for the Auditory EEG SPGC, since most submissions to the challenge were separated by only a fine margin.

\subsection{Comparison of the envelope-based decoder and the FFR-based decoder}

%% Results to comment on
% 1. Envelope-based decoders better than FFR
% -> response more strongly represented in EEG (EEG has low power at F0, difference strength of underlying response?)
% 2. Accuracies of the two decoders were not that correlated, nor were the sigmoid outputs.
% -> maybe can combine both into a better classifier; used correlation as check for shift/overfitting in SPGC.
% 3. Adverse relationship between pitch and FFR decoding accuracy (points/Hz). Smaller effect than expected considering prior work. Not significant for envelope. Future work could consider more systematically what speaker charactersitics impact speech-FFRs.

Overall, envelope-based decoders achieved much higher classification accuracies than FFR-based decoders. This is due to the fact that speech-FFRs are not that strongly represented in EEG signals, in part due to the low SNR of EEG signals at high frequencies. The accuracies of the two averaged decoders were not that correlated ($R=0.286$), nor were their sigmoid outputs ($R=0.233$). We therefore hypothesised that through combining the two decoders, we could produce a composite decoder which performs better than its constituent parts.

The composite decoder did in fact achieve higher classification accuracies than either of the two averaged decoders, suggesting that the underlying EEG responses capture different information which is relevant to the match-mismatch task. It is certainly the case that the neural processes which generate the two EEG responses expose different aspects of neural speech tracking, and the speech-FFR is relatively more robust against changes in cognitive factors such as attention. Aside from neurophysiological differences, the two responses occur at vastly different frequency scales and are most likely not affected in a similar manner by the same sources of artefacts.

The classification accuracy of the averaged FFR-based decoder varied significantly with the pitch of the speech material. This effect was smaller than expected based on the works of Kulasingham~\textit{et al.} and Puffay~\textit{et al.}~\cite{Kulasingham2020, puffay22_interspeech}. Kulasingham~\textit{et al.} found that the transfer function of the speech-FFR TRF had almost no power above around \qty{130}{\hertz}; it is possible that the nonlinear nature of our decoding approach allows for the retrieval of higher-frequency responses. Puffay~\textit{et al.} found that a spectral speech-FFR decoder, very similar to the FFR-based decoder used in this work, could not achieve statistically significant decoding accuracies for female-narrated speech material at all. Perhaps our decoder benefits from the stronger nature of the envelope-related speech-FFR. Future work should investigate what precisely is encoded by both the EEG module and the stimulus module of the FFR-based decoder.

\subsection{Composite decoders}

We combined the averaged envelope-based and FFR-based decoders using a linear classifier, LDA. The decision boundary of this classifier is overlaid on the data that was used to train it in Figure~\ref{fig:decoder-comparison}b. Clearly, the classifier assigns more weight to the predictions of the envelope-based decoder, which was the most accurate decoder. In fact, from the equation of the decision boundary, the exact weights which the classifier assigns to the sigmoid outputs of the two decoders were derived: these are $0.61$ for the envelope-based decoder, and $0.39$ for the FFR-based decoder, respectively. Despite achieving a considerably worse performance than the envelope-based decoder, the FFR-based decoder must carry a considerable amount of complementary information in order to be assigned such a high weight.

When evaluated on the heldout dataset, the composite decoder significantly outperformed the next best population-based decoder, which was the averaged envelope-based decoder. The significance of this result was not replicated for the speech-in-quiet condition of the ICL dataset: this could be due to the lower performance of the constituent FFR-based decoder on this dataset, which was presumably due to the high average pitch of the speech material (\qty{182}{Hz}). 

Puffay~\textit{et al.} also combined spectral FFR-based decoders with envelope-based decoders to solve the match-mismatch task~\cite{puffay22_interspeech}. In that study, the authors trained the two constituent decoders jointly, whereas we applied the LDA classifer post-hoc. By jointly training the decoders in the manner of Puffay~\textit{et al.}, it is possible that the composite decoder may have achieved even higher classification accuracies.

\subsection{Decoder fine tuning}

Different individuals produce EEG signals with very different characteristics. Monesi~\textit{et al.} have shown that by fine-tuning population-based decoders to individual participants, improved match-mismatch decoding accuracies can be achieved. In fact, this method may yield better results than would be achieved by training participant-specific models from scratch~\cite{monesi2020ICASSP}. In our results, a statistically significant decoding improvement was achieved by fine-tuning the population decoders to individual participants. When evaluated on the testing portion of the development dataset, the mean improvement in the decoding accuracy (taken across participants) due to fine-tuning was \qty{4.0}{\percent} for the envelope-based decoder, and \qty{2.9}{\percent} for the FFR-based decoder, when compared with the respective averaged decoders. The significance of this improvement was also replicated using the heldout dataset, although the effect size was somewhat smaller for the FFR-based decoder.

For the ICASSP Auditory EEG Decoding SPGC, we attempted to form fully-fine-tuned composite decoders, by combining the sigmoid outputs of the fine-tuned decoders using a linear classifier which was personalised to every participant. However, that submission achieved poorer results than those achieved by the fine-tuned envelope-based decoders alone; the LDA classifiers were overfitted to the small amount of data available per participant in the testing portion of the development dataset. In this work, we formed partially individualised composite decoders, by combining the sigmoid outputs of the fine-tuned decoders using the same population-based LDA classifier that was trained using the entire testing portion of the development dataset. This decoder achieved a particularly high decoding accuracy of \qty{83.79}{\percent}.

The improvement offered by fine-tuning the decoders to individual participants was not exceedingly large. In other words, the population-based decoders demonstrated a remarkable ability to generalise between participants whilst maintaining high classification accuracies. Accou~\textit{et al.} reported that the classification accuracy (as evaluated on a population of participants) of their envelope-based match-mismatch classifier reached a plateau when 28 participants were included in the training dataset, and did not increase with an increasing number of training participants~\cite{Accou2021}. Since our decoder architecture was based on that of Accou~\textit{et al.}, it is unlikely that the gap between the performance of our population and fine-tuned decoders can be closed by using a larger training dataset which includes even more participants.

\subsection{Generalisation of decoders to distinct dataset}

Finally, we evaluated our already-trained decoders on the ICL dataset, which is entirely independent of SparrKULee. The ICL dataset was completely unseen during the development and training of the decoders. All three decoders generalised remarkably well to the EEG data recorded under speech-in-quiet conditions in this dataset, even when the speech was  in a language that the participants did not understand.

When speech was played in the presence of background noise, the match-mismatch classification accuracy of the envelope-based decoder deteriorated. This is to be expected, since cortical envelope tracking is known to change during speech-in-noise perception, and moreover such low-SNR listening conditions were not represented in the training dataset. The fact that the performance of the envelope-based decoder did not deteriorate in the foreign-language speech-in-quiet condition suggests that the performance of the envelope-based decoder is predominantly affected by speech clarity, rather than by speech comprehension. Etard~\textit{et al.} have previously shown that both the clarity and comprehension of speech may be differentially decoded from EEG recordings, by using linear models to assess neural envelope tracking~\cite{Etard2019}. The performance of the FFR-based decoder was rather consistent across all of the listening conditions, dropping only slightly in the lowest-SNR conditions.

The ICL dataset also included EEG recorded under competing-speakers conditions. When applied in the usual match-mismatch setting, the envelope-based decoder performed better for the attended speaker than the ignored speaker, presumably reflecting how cortical responses to the speech envelope are modulated by selective auditory attention. Indeed, by drawing matched segments from the attended speech stream, and temporally aligned `mismatched' segments from the ignored speech stream, we showed that the envelope-based decoder can be used as an auditory attention decoder which achieved a mean accuracy of 62.86\%. When using the FFR-based decoder, there was no significant difference between the match-mismatch decoding accuracies of the attended speaker and the ignored speaker. This decoder achieved a  mean attention decoding accuracy of 52.08\%, which was significantly greater than 50\%, suggesting a subtle attentional effect. Since the effect was not strong, the decoding accuracies for individuals fell mostly within the 95\% confidence interval of a random binary classifier, and the composite decoder did not outperform the envelope-based decoder at the auditory attention decoding task. 

We also used the ICL dataset to investigate how the length of the EEG and speech-feature segments affects the decoding accuracy. Although the decoders were trained using segments of \qty{3}{s} in duration, they can be evaluated using segments of any duration. It is shown in Table~\ref{table:seglen} that the match-mismatch decoding accuracy reliably increases for all decoders when the segment length is increased to \qty{5}{s} or \qty{10}{s}. For the FFR-based decoder, the attention decoding accuracy did not increase when the segment length increased.

There were some differences between the experimental setups of the ICL dataset and SparrKULee, and several EEG channels which were present in SparrKULee were missing from the ICL dataset and required interpolation. For these reasons, the finding that the decoders generalised so well between the two datasets is particularly remarkable. That the match-mismatch decoders could also serve as auditory attention decoders was also an important finding. Usually, attention decoders are developed using relatively small EEG datasets consisting of data from participants who listened to competing speakers. Our results show that these datasets may be supplemented by other datasets in which participants listened to speech material under various other listening conditions; these other datasets are more numerous and, typically, contain more data than the competing-speakers datasets. The finding that these datasets are compatible therefore opens new avenues for learning from vast amounts of auditory EEG data.

\section*{ACKNOWLEDGMENT}
Michael Thornton is supported by the UKRI CDT in AI for Healthcare \href{http://ai4health.io}{http://ai4health.io} (Grant No. P/S023283/1)

\section*{Code availability}
Supporting Python code is available at \href{https://github.com/Mike-boop/match-mismatch-decoders-ojsp-2023}{https://github.com/Mike-boop/match-mismatch-decoders-ojsp-2023}. This package contains all the functions used for data preprocessing, model training, and analysis.

\bibliographystyle{IEEEbib}
\bibliography{bibliography}

\begin{IEEEbiography}
% \begin{IEEEbiography}[{\includegraphics[width=1in,height=1.25in,clip,keepaspectratio]{a1.png}}]
{MIKE D. THORNTON}~photograph and biography not available
at the time of publication.
% ~and all authors may include biographies.
% Biographies are often not included in conference-related papers. This author
% is an IEEE Fellow. The first paragraph may contain a place and/or date of
% birth (list place, then date). Next, the author's educational background is
% listed. The degrees should be listed with type of degree in what field,
% which institution, city, state, and country, and year the degree was earned.
% The author's major field of study should be lower-cased.

% The second paragraph uses the pronoun of the person (he or she) and not the
% author's last name. It lists military and work experience, including summer
% and fellowship jobs. Job titles are capitalized. The current job must have a
% location; previous positions may be listed without one. Information
% concerning previous publications may be included. Try not to list more than
% three books or published articles. The format for listing publishers of a
% book within the biography is: title of book (publisher name, year) similar
% to a reference. Current and previous research interests end the paragraph.

% The third paragraph begins with the author's title and last name
% (e.g., Dr.\ Smith, Prof.\ Jones, Mr.\ Kajor, Ms.\ Hunter). List any memberships in
% professional societies other than the IEEE. Finally, list any awards and
% work for IEEE committees and publications. If a photograph is provided, it
% should be of good quality, and professional-looking.
\end{IEEEbiography}

\begin{IEEEbiographynophoto}
{DANILO P. MANDIC}~(Fellow, IEEE)~photograph and biography not available at the time
of publication.
\end{IEEEbiographynophoto}

\begin{IEEEbiographynophoto}
{TOBIAS J. REICHENBACH}~photograph and biography not available
at the time of publication.
\end{IEEEbiographynophoto}

\vfill\pagebreak

\end{document}